\documentclass[%
 aip,
 amsmath,amssymb,
reprint,%
]{revtex4-1}

\usepackage{graphicx}
\usepackage{dcolumn}
\usepackage{bm}

\usepackage[utf8]{inputenc}
\usepackage[T1]{fontenc}
\usepackage{mathptmx}
\usepackage{etoolbox}
\usepackage{feynmp-auto}
\usepackage{braket}
\usepackage{framed}
\usepackage{caption}
\usepackage{simplewick}
\usepackage{tikz}
\usepackage{bbold}
\usepackage{gensymb}
\usetikzlibrary{patterns}
\usetikzlibrary{shapes.geometric}
\usepackage{pgfplots}
\usetikzlibrary{perspective,3d,patterns.meta}

\makeatletter
\def\@email#1#2{%
 \endgroup
 \patchcmd{\titleblock@produce}
  {\frontmatter@RRAPformat}
  {\frontmatter@RRAPformat{\produce@RRAP{*#1\href{mailto:#2}{#2}}}\frontmatter@RRAPformat}
  {}{}
}%
\makeatother

\def\fmfhline#1{\fmfcmd{input hline; hline (\fmfpfx{#1});}}

\def\fmfdhline#1{\fmfcmd{input dhline; dhline (\fmfpfx{#1});}}

\begin{document}

\preprint{AIP/123-QED}

\title[Diagrams and symmetry in polaritonic coupled cluster theory]{Diagrams and symmetry in polaritonic coupled cluster theory}
\author{Laurenz Monzel}
 \affiliation{Fachrichtung Chemie, Universität des Saarlandes, Campus B2.2, D-66123 Saarbrücken, Germany}
\author{Stella Stopkowicz}
\affiliation{Fachrichtung Chemie, Universität des Saarlandes, Campus B2.2, D-66123 Saarbrücken, Germany}%
\affiliation{Hylleraas Centre for Quantum Molecular Sciences, Department of Chemistry, University of Oslo,
P.O. Box 1033 Blindern, N-0315 Oslo, Norway}
\date{\today}

\begin{abstract}
   We present a diagrammatic notation to derive the quantum-electrodynamic coupled cluster (QED-CC) equations 
   needed for the description for polaritonic ground and excited states. 
   Our presented notation is a generalization of the existing diagrammatic notation of standard electronic coupled-cluster theory.
   Further, we discuss point-group symmetry inside a cavity and its exploitation within a QED-CC code for efficiency and the targeted treatment of excited states. 
\end{abstract}

\maketitle

\section{Introduction}
Molecular polaritons which are hybrid light-matter states can be generated by placing molecules in small cavities.
\cite{daskalakis2014nonlinear,thomas2016ground,basov2016polaritons}
By carefully choosing the frequency of the cavity, molecular processes can be activated, inhibited, modified and controlled.\cite{ebbesen2016hybrid}
A molecular polaritonic state can be viewed as 
a quasiparticle formed by photons and electronic excitations.
These hybrid states gain importance in the so-called strong-coupling regime,\cite{flick2018strong,schaefer2018abinitio} where the coupling between light and matter strongly influences  characteristic properties of the system. The coupling strength increases by reduction of the cavity size as well as a higher number of molecules (emitters).
The possibility to influence molecular systems in a non-invasive manner motivated much research on both the experimental\cite{ebbesen2016hybrid,ebbesen2021chemistry, chikkaraddy2016single,benson2011assembly,stranius2018selective,garcia2021manipulating} as well as the theoretical sides.\cite{ruggenthaler2023understanding, ruggenthaler2014quantum, haugland2020coupled,castagnola2024polaritonic}

Oftentimes the description of the light-matter interaction is based on effective model Hamiltonians, e.g., the Dicke, Jaynes-Cummings  or Tavis-Cumming model. Such models are typically useful for a qualitative description of the interaction. However, if one aims at a quantitative description of molecular polaritonic states from first principles a more rigorous formulation is required.  
Most of the model Hamiltonians reduce the broad energetic landscape of the matter system to a two-level system which can be treated to a large extend analytically, but at the cost of loosing the response of the matter system.
To overcome this limitation, self-consistent field methods as QED Hartree-Fock (QED-HF)\cite{haugland2020coupled}, QED CAS-SCF\cite{vu2024cas}, and, in particular, QED density-functional theory (QEDFT)\cite{ruggenthaler2014quantum,schafer2021making} have been developed.
Therefore, by also including the electronic degrees of freedom, such methods are able to give a more comprehensive picture of the system.   
While QEDFT is able to describe also electron correlation, it comes with similar drawbacks as standard electronic DFT theory, as the exact electron-electron exchange correlation functional and also the electron-photon exchange correlation potential are not known. There is hence an ongoing intensive research on the formulation of suitable functionals. \cite{lu2024electron}
For a even more rigorous description of electron and electron-photon correlation within wave-function-based methods, perturbation theory (QED-PT),\cite{bauer2023perturbation} full configuration interaction theory (QED-FCI) \cite{mordovina2020polaritonic,haugland2021intermolecular} as well as coupled cluster theory (QED-CC)\cite{haugland2020coupled} have been formulated. 
While PT may converge only slowly (or not at all) towards the exact limit and FCI is impractical for more than just the smallest benchmark systems, it is mostly CC and its variants which can be seen as the method of choice when highly accurate results are needed.
In 2020, the first QED-CC implementations were presented\cite{mordovina2020polaritonic,haugland2020coupled} and 
various truncation schemes for the cluster operator were explored. 
The ansatz introduced by Haugland et al.\cite{haugland2020coupled} has also been adopted by the group of DePrince for the calculation of electron attachment processes using QED-EOM-EA.\cite{deprince2022equation}
Riso et al. showed that the same truncation scheme can be applied for circular polarized cavities in the minimal-coupling picture.\cite{riso2023strong}

Since the CC equations are quite complicated, diagrammatic techniques for their derivation have turned out very useful. In particular, 
the diagrammatic notation by Kucharsky and Bartlett is able to represent all terms in a compact manner 
\cite{kucharski1986fifth} and has since advanced to the standard way of dealing with diagrams in CC theory. \cite{crawford2007introduction,shavitt2009many}
In this work we present a diagrammatic notation and general rules for the derivation of the QED-CC equations. 
This notation is a generalization of the established diagrammatic notation of electronic CC theory and allows to avoid otherwise lengthy mathematical expressions.
We will give all needed equations for the QED-CCSD-1-SD and QED-CCSD-12-SD truncation schemes which have first been published by Haugland et al.\cite{haugland2020coupled} and Philbin et al.\cite{philbin2023molecular}, respectively.
The equations will be derived for a general cavity including an arbitrary number of modes. 

Further, we discuss point-group symmetry of molecules in cavities and its exploitation in the speed-up of the respective calculations as well as the targeting of polaritonic states. 
An efficient implementation of point-group symmetry in CC calculations based on the direct-product decomposition was put forward by Stanton, Gauss, Watts, and Bartlett\cite{stanton1991symmetry} which is adapted here for the use in our QED-CC implementation. 

In section (\ref{sec:polham}) we introduce the polaritonic Hamiltonian in the dipole approximation. 
Section (\ref{sec:diagramms})  presents how the diagrammatic notation for electronic CC theory can be generalized in order to also represent contractions over photonic operators.
The section concludes with the necessary equations for the implementations of the CCSD-1-SD and CCSD-12-SD truncations.
In section (\ref{sec:eom}) we present the QED-EOM-CC equations and give all needed equations in the CCSD-1-SD and CCSD-12-SD truncation schemes. We also discuss the modification of a modified configuration interaction singles (CIS) guess to be used for the generation of initial guess vectors for the QED-EOM-CC calculation.
Section (\ref{sec:symmetry}) discusses how point-group symmetry changes inside a cavity and how it can be employed in QED-(EOM)-CC calculations. 
Further, we discuss in this section the coupling mechanism of excited states and in particular which states couple and give rise to the typical Rabi splittings.
In section (\ref{sec:calculations}) we then discuss polariton formation for the H$_2$  and H$_2^-$ molecules. 
These rather simple molecules already show how challenging the interpretation of the polaritonic spectra can get and that the truncation level of the cluster operator must be chosen carefully.

\section{Polaritonic Hamiltonian}
\label{sec:polham}
In its most general form, the Hamiltonian of a molecule in a cavity is given by 
\begin{equation}
   \hat{H} = \hat{H}_\text{el} + \hat{H}_\text{int} + \hat{H}_\text{ph}\;,
   \label{eq:minham}
\end{equation}
where $\hat{H}_\text{el}$ and  $\hat{H}_\text{ph}$ describes the isolated electronic and photonic part  and $\hat{H}_\text{int}$ describes their interaction.
We start from the Pauli-Fierz Hamiltonian in the dipole approximation in the polaritonic Born-Oppenheimer approximation\cite{haugland2020coupled}
\begin{equation}
	\begin{split}
      \hat{H} &= \sum_{pq} \tilde{h}_{pq} \hat{p}^\dagger \hat{q} + \frac{1}{4}\sum_{pqrs} \tilde{g}_{pqrs} \hat{p}^\dagger \hat{q}^\dagger \hat{s} \hat{r}\;. \\
      & + \sum_{pq\alpha} \tilde{d}_{pq}^\alpha  (\hat{\alpha}^\dagger + \hat{\alpha}^\dagger) \hat{p}^\dagger \hat{q} + \sum_{\alpha} \omega_\alpha  \hat{\alpha}^\dagger  \hat{\alpha}	\;.
	\end{split}
   \label{eq:ham}
\end{equation}
For the creation and annihilation operators we use the notation that electronic operators are written in Latin letters $\hat{p}^\dagger$ and $\hat{p}$ while the photonic operators are written using Greek letters $\hat{\alpha}^\dagger$ and $\hat{\alpha}$.
The Hamiltonian in (\ref{eq:ham}) is written in a coherent-state basis,\cite{haugland2020coupled} where the first term corresponds to the modified one-electron integrals $\tilde{h}_{pq}$ which contain contributions from the dipole self-energy
\begin{equation}
      \tilde{h}_{pq} =  h_{pq} + \sum_{\alpha}\sum_r \frac{ \tilde{d}_{pr}^\alpha \tilde{d}_{rq}^\alpha}{\omega_\alpha} 		\;.
	\label{eq:1e_int}
\end{equation}
The one-electron integrals $h_{pq}$ include the electrostatic electron-nuclei interaction as well as the kinetic energy of the electrons.
Here, $\tilde{d}_{pq}$ are the dipole integrals in the coherent state basis which are given as
\begin{equation}
   \tilde{d}_{pq}^\alpha = \sqrt{\frac{\omega_\alpha}{2}} \boldsymbol{\lambda}_\alpha \cdot (\boldsymbol{d}_{pq} - \delta_{pq} \braket{\boldsymbol{d}} ) \;,
	\label{}
\end{equation}
with $\boldsymbol{d}_{pq}$ as standard dipole integrals (including both the electronic and nuclear contribution).
The vector $\boldsymbol{\lambda}_\alpha$ defines the orientation of the electric field in the cavity and is composed of a unit vector scaled by the coupling strength $\boldsymbol{\lambda}_\alpha = \boldsymbol{\epsilon}_\alpha \lambda_\alpha $ both of which are parameters in the calculation. The coupling strength controls the strength of the interaction between the electronic and photonic parts.
The coupling strength connected the quantization volume $V_\alpha$ of the cavity
\begin{equation}
    \lambda_\alpha = \frac{1}{\sqrt{\epsilon_0 V_\alpha }}
    \label{eq:couplingstrength}
\end{equation}
where $\epsilon_0$ is the vacuum permittivity.
For a general cavity, the $\alpha$ indices run over all frequencies and polarization vectors in the system characterizing the electromagnetic (EM) modes of the cavity.
The modified two-electron integrals $\tilde{g}_{pqrs}$ also contain contributions from the dipole self-energy and are given as
\begin{equation}
   \tilde{g}_{pqrs} = \braket{pq||rs} + \sum_{\alpha} \frac{ \tilde{d}_{pr}^\alpha \tilde{d}_{qs}^\alpha - \tilde{d}_{ps}^\alpha \tilde{d}_{qr}^\alpha}{\omega_\alpha}.
	\label{eq:2e_int}
\end{equation}
The third term in (\ref{eq:ham}) is the bilinear electron-photon coupling operator 
\begin{equation}
	\begin{split}
      \hat{H}_\text{bil} = \sum_{pq\alpha} \tilde{d}_{pq}^\alpha  (\hat{\alpha}^\dagger + \hat{\alpha}^\dagger) \hat{p}^\dagger \hat{q} .
      \label{}
	\end{split}
   \label{eq:bil}
\end{equation}
A mean-field solution for the polaritonic ground-state $\ket{0,0}$ can be found by setting up an appropriate Fock operator and solving for the coherent-state Hartree-Fock equations, as presented by Haugland et al.\cite{haugland2020coupled}
The first entry in $\ket{0,0}$ refers to the electronic state (0 for ground-state) and the second entry corresponds to the number of photons (here zero).
The state $\ket{0,0}$ serves as reference state for the coupled-cluster wave function.

\section{Diagrammatic notation for QED-Coupled Cluster theory}
\label{sec:diagramms}
In order to obtain a graphical way for the derivation of the  QED-CC equations we are introducing in this chapter a generalization of the well established diagrammatic notation for CC theory.\cite{kucharski1986fifth,crawford2007introduction,shavitt2009many}
We assume that the reader is familiar with the diagrammatic notation and will discuss here only terms and concepts that go beyond standard CC diagrams. For an introduction to CC diagrams, see for example refs \onlinecite{shavitt2009many} and \onlinecite{crawford2007introduction}.
In the QED-CC framework, a new type of "line" is introduced for the creation and annihilation of photons.
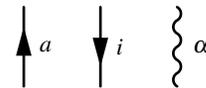
\begin{figure}[h!]
\begin{fmffile}{particle-line}
 \begin{fmfgraph*}(20,30)
 \fmfstraight
\fmftop{t1}
\fmfbottom{b1}
\fmf{fermion,label=$a$}{b1,t1}
 \end{fmfgraph*}
 \end{fmffile}
\begin{fmffile}{holw-line}
 \begin{fmfgraph*}(20,30)
 \fmfstraight
\fmftop{t1}
\fmfbottom{b1}
\fmf{fermion,label=$i$,l.s=left}{t1,b1}
 \end{fmfgraph*}
 \end{fmffile}
\begin{fmffile}{photon-line}
 \begin{fmfgraph*}(20,30)
 \fmfstraight
\fmftop{t1}
\fmfbottom{b1}
\fmf{photon,label=$\alpha$}{b1,t1}
 \end{fmfgraph*}
 \end{fmffile}
 \label{fig:elementar-lines}
 \caption{Line types for particles, holes, photons appearing in QED-CC theory. }
 \label{fig:lines}
\end{figure}\\
For the electrons, the Fermi-vacuum (the HF-determinant) is used while the photonic vacuum  corresponds to the physical vacuum. 
In contrast to the electrons, therefore, no reformulation in particles and holes is necessary and the upwards pointing arrow for photonic lines (wavy line in Fig.  \ref{fig:lines}) can therefore be omitted.
Operators can then be set up in a familiar manner, where lines facing upwards are referring to a creation of the respective quasiparticle ($\hat{a}^\dagger,\hat{i} \text{ and } \hat{\alpha}^\dagger$) and lines facing downwards to their annihilation ($\hat{a},\hat{i}^\dagger \text{ and } \hat{\alpha}$).
Here and throughout this work, $i,j,k,\cdots$ are referring to occupied orbitals, $a,b,c,\cdots$ to virtual orbitals and $p, q ,r,\cdots$ to a general orbital.
In the dipole approximation, the normal-ordered Hamiltonian $\hat{H}_\text{N}$ is defined as 
\begin{equation}
   \hat{H}_N = \hat{H} - \braket{0,0|\hat{H}|0,0}
   \label{eq:normal_ham}
\end{equation}
and contains, besides the bare electronic operators two operators that include photonic indices.
The normal ordered electronic operator is 
\begin{equation}
   (\hat{H}_\text{el})_N = \sum_{pq} \tilde{f}_{pq} \{\hat{p}\hat{q}^\dagger\} + \frac{1}{4}\sum_{pqrs} \tilde{g}_{pqrs} \{\hat{p}\hat{q}\hat{s}^\dagger \hat{r}^\dagger \}\;,
   \label{}
\end{equation}
with the Fock operator in the coherent state basis 
\begin{equation}
   \tilde{f}_{pq} = \tilde{h}_{pq} + \sum_{i} \tilde{g}_{piqi}\;.
   \label{eq:fock}
\end{equation}
The braces $\{\cdots\}$ indicate the normal ordering of the creation and anihilation operators with respect to the Fermi vacuum.
The operators containing  photonic indices are the operator for the EM field $\hat{H}_\text{ph}$ as well as the bilinear coupling operator $\hat{H}_\text{bil}$. 
$\hat{H}_\text{ph}$ contains only photonic indices and can be represented diagrammatically as
\begin{equation}
   \hat{H}_\text{ph}=\sum_\alpha \omega_{\alpha} \hat{\alpha}^\dagger\alpha \leftrightarrow 
 \begin{minipage}{0.05\textwidth}
\begin{fmffile}{EM-field}
 \begin{fmfgraph*}(20,30)
 \fmfstraight
\fmftop{t1}
\fmfbottom{b1}
\fmf{phantom}{b1,m1,t1}
\fmf{photon}{b1,t1}
\fmfv{decor.shape=square,decor.filled=empty,decor.size=3thick}{m1}
 \end{fmfgraph*}
 \end{fmffile}
 \end{minipage}\;.
 \label{eq:em-field}
\end{equation}
As usual, a downwards facing line represents the annihilation and an upwards facing line the creation of the respective quasiparticle.
This operator is number conserving and is already in normal-ordered form by construction.\\
$\hat{H}_\text{bil}$ may 
  be expressed as
\begin{equation}
	\begin{split}
      & \hat{H}_\text{bil} = \sum_{pq}^\alpha \tilde{d}_{pq}^\alpha  (\hat{\alpha}^\dagger + \hat{\alpha} ) \{\hat{p}^\dagger \hat{q}\} + \sum_{i \alpha} \tilde{d}_{ii}^\alpha (\hat{\alpha}^\dagger + \hat{\alpha} )\\
	 & \leftrightarrow
    \begin{minipage}{0.9\textwidth}
       \begin{minipage}{0.10\textwidth}
       \centering
       \begin{fmffile}{1e1p-ham6}
        \begin{fmfgraph*}(20,30)
        \fmfstraight
       \fmftop{t1,t2}
       \fmfbottom{b1,b2}
       \fmf{fermion}{b1,m1,t1}
       \fmf{phantom}{b2,m2,t2}
       \fmf{photon,tension=.0}{t2,m2}
       \fmf{dashes,tension=.0}{m1,m2}
       \fmfdot{m1,m2}
        \end{fmfgraph*}
        \end{fmffile}
        \end{minipage}
        +
        \begin{minipage}{0.10\textwidth}
        \centering
       \begin{fmffile}{1e1p-ham7}
        \begin{fmfgraph*}(20,30)
        \fmfstraight
       \fmftop{t1,t2}
       \fmfbottom{b1,b2}
       \fmf{fermion}{t1,m1,b1}
       \fmf{phantom}{b2,m2,t2}
       \fmf{photon,tension=.0}{t2,m2}
       \fmf{dashes,tension=.0}{m1,m2}
       \fmfdot{m1,m2}
        \end{fmfgraph*}
        \end{fmffile}
        \end{minipage}
        +
         \begin{minipage}{0.10\textwidth}
         \centering
       \begin{fmffile}{1e1p-ham8}
        \begin{fmfgraph*}(20,30)
        \fmfstraight
       \fmftop{t1,t2}
       \fmfbottom{b1,b2}
       \fmf{phantom}{t1,m1,b1}
       \fmffreeze
       \fmf{phantom,tension=.0}{m1,b1}
       \fmfi{fermion}{vpath (__b1,__m1) rotatedaround(vloc(__m1),-19)}
       \fmf{phantom,tension=.0}{b1,m1}
       \fmfi{fermion}{vpath (__b1,__m1) rotatedaround(vloc(__m1),19)}
       \fmf{phantom}{b2,m2,t2}
       \fmf{photon,tension=.0}{t2,m2}
       \fmf{dashes,tension=.0}{m1,m2}
       \fmfdot{m1,m2}
        \end{fmfgraph*}
        \end{fmffile}
        \end{minipage}
       +
         \begin{minipage}{0.10\textwidth}
         \centering
       \begin{fmffile}{1e1p-ham9}
        \begin{fmfgraph*}(20,30)
        \fmfstraight
       \fmftop{t1,t2}
       \fmfbottom{b1,b2}
       \fmf{phantom}{t1,m1,b1}
       \fmffreeze
       \fmf{phantom,tension=.0}{m1,t1}
       \fmfi{fermion}{vpath (__t1,__m1) rotatedaround(vloc(__m1),-19)}
       \fmf{phantom,tension=.0}{t1,m1}
       \fmfi{fermion}{vpath (__m1,__t1) rotatedaround(vloc(__m1),19)}
       \fmf{phantom}{b2,m2,t2}
       \fmf{photon,tension=.0}{t2,m2}
       \fmf{dashes,tension=.0}{m1,m2}
       \fmfdot{m1,m2}
        \end{fmfgraph*}
        \end{fmffile}
        \end{minipage}
    \end{minipage} \\
	 & + 
    \begin{minipage}{0.9\textwidth}
       \begin{minipage}{0.10\textwidth}
       \centering
       \begin{fmffile}{1e1p-ham1}
        \begin{fmfgraph*}(20,30)
        \fmfstraight
       \fmftop{t1,t2}
       \fmfbottom{b1,b2}
       \fmf{fermion}{b1,m1,t1}
       \fmf{phantom}{b2,m2,t2}
       \fmf{photon,tension=.0}{b2,m2}
       \fmf{dashes,tension=.0}{m1,m2}
       \fmfdot{m1,m2}
        \end{fmfgraph*}
        \end{fmffile}
        \end{minipage}
        +
        \begin{minipage}{0.10\textwidth}
        \centering
       \begin{fmffile}{1e1p-ham2}
        \begin{fmfgraph*}(20,30)
        \fmfstraight
       \fmftop{t1,t2}
       \fmfbottom{b1,b2}
       \fmf{fermion}{t1,m1,b1}
       \fmf{phantom}{b2,m2,t2}
       \fmf{photon,tension=.0}{b2,m2}
       \fmf{dashes,tension=.0}{m1,m2}
       \fmfdot{m1,m2}
        \end{fmfgraph*}
        \end{fmffile}
        \end{minipage}
        +
         \begin{minipage}{0.10\textwidth}
         \centering
       \begin{fmffile}{1e1p-ham3}
        \begin{fmfgraph*}(20,30)
        \fmfstraight
       \fmftop{t1,t2}
       \fmfbottom{b1,b2}
       \fmf{phantom}{t1,m1,b1}
       \fmffreeze
       \fmf{phantom,tension=.0}{m1,b1}
       \fmfi{fermion}{vpath (__b1,__m1) rotatedaround(vloc(__m1),-19)}
       \fmf{phantom,tension=.0}{b1,m1}
       \fmfi{fermion}{vpath (__b1,__m1) rotatedaround(vloc(__m1),19)}
       \fmf{phantom}{b2,m2,t2}
       \fmf{photon,tension=.0}{b2,m2}
       \fmf{dashes,tension=.0}{m1,m2}
       \fmfdot{m1,m2}
        \end{fmfgraph*}
        \end{fmffile}
        \end{minipage}
         +
         \begin{minipage}{0.10\textwidth}
         \centering
       \begin{fmffile}{1e1p-ham4}
        \begin{fmfgraph*}(20,30)
        \fmfstraight
       \fmftop{t1,t2}
       \fmfbottom{b1,b2}
       \fmf{phantom}{t1,m1,b1}
       \fmffreeze
       \fmf{phantom,tension=.0}{m1,t1}
       \fmfi{fermion}{vpath (__t1,__m1) rotatedaround(vloc(__m1),-19)}
       \fmf{phantom,tension=.0}{t1,m1}
       \fmfi{fermion}{vpath (__m1,__t1) rotatedaround(vloc(__m1),19)}
       \fmf{phantom}{b2,m2,t2}
       \fmf{photon,tension=.0}{b2,m2}
       \fmf{dashes,tension=.0}{m1,m2}
       \fmfdot{m1,m2}
        \end{fmfgraph*}
        \end{fmffile}
        \end{minipage}
     \end{minipage}\\
	 & + 
    \begin{minipage}{0.9\textwidth}
      \begin{minipage}{0.1\linewidth}
      \centering
      \begin{fmffile}{1e1p-ham10}
      \begin{fmfgraph*}(15,25)
      \fmfstraight
      \fmftop{t1,t2,t3}
      \fmfbottom{b1,b2,b3}
      \fmf{phantom}{b1,m1,t1}
      \fmf{phantom}{b2,m2,t2}
      \fmf{phantom}{b3,m3,t3}
      \fmf{photon,tension=.0}{b1,m1}
      \fmf{dashes,tension=.0}{m1,m2}
      \fmffreeze
      \fmfshift{(4.6,0)}{m3}
      \fmf{fermion,right}{m3,m2,m3}
      \fmfdot{m1,m2}
      \end{fmfgraph*}
      \end{fmffile}
      \end{minipage}
		+
      \begin{minipage}{0.1\linewidth}
      \centering
      \begin{fmffile}{1e1p-ham11}
      \begin{fmfgraph*}(15,25)
      \fmfstraight
      \fmftop{t1,t2,t3}
      \fmfbottom{b1,b2,b3}
      \fmf{phantom}{b1,m1,t1}
      \fmf{phantom}{b2,m2,t2}
      \fmf{phantom}{b3,m3,t3}
      \fmf{photon,tension=.0}{t1,m1}
      \fmf{dashes,tension=.0}{m1,m2}
      \fmffreeze
      \fmfshift{(4.6,0)}{m3}
      \fmf{fermion,right}{m3,m2,m3}
      \fmfdot{m1,m2}
      \end{fmfgraph*}
      \end{fmffile}
      \end{minipage}
    \end{minipage} 
  \end{split}\;,
  \label{eq:bil-op}
  .
\end{equation}
This operator is not number conserving as either a photon is created (upward-facing photon line) or annihilated (downward-facing photon line).
Also, this operator is not in normal order, resulting in the appearance of the so-called self-contractions, where the operator is contracted with itself - see the last two diagrams which correspond to the second term in the algebraic equation above.
These diagrams are also called bubble diagrams\cite{shavitt2009many} and are obtained by reordering the full operator in normal order using Wick's theorem\cite{wick1950}
\begin{equation}
   \begin{split}
   \hat{\alpha}\hat{p}^\dagger\hat{q} & = \hat{\alpha}\{\hat{p}^\dagger\hat{q} \} + \hat{\alpha} \contraction{}{p}{q}{} 
   \hat{p}^\dagger \hat{q}\\
   & = \hat{\alpha}\{\hat{p}^\dagger\hat{q} \} + \hat{\alpha} n_p \delta_{pq}\;.
   \end{split}
   \label{}
\end{equation}
where $n_p$ is the occupation number of the $p$'th orbital of the reference state $\ket{0,0}$.
However, note that all bubble contractions vanish for a coherent-state HF reference wave function since the dipole expectation value is subtracted from the dipole operator:  
\begin{equation}
  \begin{minipage}{0.2\linewidth}
 \centering
      \begin{fmffile}{bubble}
       \begin{fmfgraph*}(25,25)
      \fmfstraight
      \fmftop{t1,t2,t3}
      \fmfbottom{b1,b2,b3}
      \fmf{phantom}{b1,m1,t1}
      \fmf{phantom}{b2,m2,t2}
      \fmf{phantom}{b3,m3,t3}
      \fmf{photon,tension=.0}{b1,m1}
      \fmf{dashes,tension=.0}{m1,m2}
      \fmffreeze
       \fmfshift{(4.6,0)}{m3}
      \fmf{fermion,left}{m3,m2,m3}
      \fmfdot{m1,m2}
       \end{fmfgraph*}
      \end{fmffile}
       \end{minipage}
       \leftrightarrow \sum_{i} \tilde{d}_{ii\alpha}  = \braket{\hat{d}_\alpha -  \braket{d_{\alpha}}} \hat{\alpha} = 0\;,
	\label{}
\end{equation}
and similar for the photon creation.
Therefore, when the molecule shows no dipole moment or when using a coherent-state HF reference state, there is no need to include bubble contractions.
Here, we list them in order provide a more general set of contractions. 

In QED-CC theory the wavefunction is parameterized as
\begin{equation}
   \ket{\Psi_\text{CC}} = \text{e}^{\hat{Q}} \ket{\text{0,0}}
	\label{}
\end{equation}
with the reference state $\ket{\text{0,0}}$. 
In the CCSD-1-SD scheme the cluster operator $\hat{Q}$ is truncated as 
\begin{equation}
	\hat{Q} = \hat{T}_1 + \hat{T}_2  +  \hat{S}_1^1 + \hat{S}_2^1 + \hat{\Gamma}_1\;
	\label{}
\end{equation}
where the mixed- and photonic cluster operators $\hat{S}$ and $\hat{\Gamma}$ can be written in second quantization and diagrammatic notation respectively,  as  
\begin{align}
	 \hat{S}_1^1 &= \sum_{ai\alpha} s_{ai}^\alpha \hat{\alpha}^\dagger\hat{a}^\dagger \hat{i}  &\leftrightarrow
       \begin{minipage}{0.30\linewidth}
       \centering
       \begin{fmffile}{S1-operator}
        \begin{fmfgraph*}(15,30)
        \fmfstraight
		  \fmftop{t1,t2}
		  \fmfbottom{b1,b2}
		    \fmf{phantom}{b1,t1}
			 \fmfi{fermion}{vpath (__b1,__t1) rotatedaround(vloc(__b1),-13)}
		    \fmf{phantom}{t1,b1}
			 \fmfi{fermion}{vpath (__t1,__b1) rotatedaround(vloc(__b1),13)}
			 \fmf{plain}{b1,b2}
			 \fmf{photon}{b2,t2}
        \end{fmfgraph*}
        \end{fmffile}
	  \end{minipage}\;,\\
	  \hat{S}_2^1 &= \frac{1}{4} \sum_{abij\alpha} s_{abij}^\alpha \hat{\alpha}^\dagger\hat{a}^\dagger\hat{b}^\dagger \hat{j}\hat{i}  &\leftrightarrow 
       \begin{minipage}{0.30\linewidth}
       \centering
       \begin{fmffile}{S2-operator}
        \begin{fmfgraph*}(30,30)
        \fmfstraight
		  \fmftop{t1,t2,t3}
		  \fmfbottom{b1,b2,b3}
		    \fmf{phantom}{b1,t1}
			 \fmfi{fermion}{vpath (__b1,__t1) rotatedaround(vloc(__b1),-13)}
		    \fmf{phantom}{t1,b1}
			 \fmfi{fermion}{vpath (__t1,__b1) rotatedaround(vloc(__b1),13)}
		    \fmf{phantom}{b2,t2}
			 \fmfi{fermion}{vpath (__b2,__t2) rotatedaround(vloc(__b2),-13)}
		    \fmf{phantom}{t2,b2}
			 \fmfi{fermion}{vpath (__t2,__b2) rotatedaround(vloc(__b2),13)}
			 \fmf{plain}{b1,b3}
			 \fmf{photon}{b3,t3}
        \end{fmfgraph*}
        \end{fmffile}
	  \end{minipage}\;,\\
		  \hat{\Gamma}_1 &= \sum_{\alpha} \gamma^\alpha \hat{\alpha}^\dagger &\leftrightarrow
       \begin{minipage}{0.30\linewidth}
       \centering
       \begin{fmffile}{G1-operator}
        \begin{fmfgraph*}(10,30)
        \fmfstraight
		  \fmftop{t1,t2,t3}
		  \fmfbottom{b1,b2,b3}
		    \fmf{photon}{b1,t1}
			 \fmfhline{b1}
        \end{fmfgraph*}
        \end{fmffile}
	  \end{minipage}\;.
	  \label{eq:cluster-op}
\end{align}

\begin{figure}[h!]
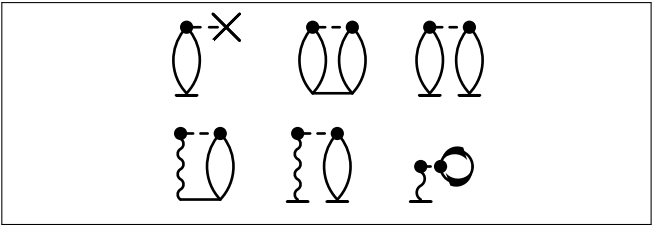

\begin{framed}
\include{include/cc_energy.tex}
\end{framed}
\caption{QED-CC energy in the dipole approximation.}
\label{fig:energy}
\end{figure}

Photonic indices will be written in the amplitudes as a superscripts to distinguish them from the electronic indices which are written as subscripts.
The amplitudes $t_{ia}$, $t_{ijab}$, $s_{ia}^\alpha$, $s_{ijab}^\alpha$ and $\gamma^\alpha$ are the target object in a CC calculation and are obtained by a set of projected equations 
\begin{equation}
	\begin{split}
      0 & = \bra{\mu,\nu}\text{e}^{-\hat{Q}} \hat{H}_\text{N} \text{e}^{\hat{Q}}\ket{0,0} \\
	0 & = \bra{\mu,\nu} \hat{H}_\text{N} + [\hat{H}_\text{N} ,\hat{Q}] + \frac{1}{2}[[\hat{H}_\text{N} ,\hat{Q}],\hat{Q}] \\
   &\quad  + \frac{1}{3!}[[[\hat{H}_\text{N} ,\hat{Q}],\hat{Q}],\hat{Q}]  + \frac{1}{4!}[[[[\hat{H}_\text{N} ,\hat{Q}],\hat{Q}],\hat{Q}],\hat{Q}]  \ket{0,0} \;.\\
\end{split}
	\label{eq:cc-amplitudes}
\end{equation}
The state $\ket{\mu,\nu}$ represents an excited state, where either the electronic or photonic or both spaces are excited.
The Baker-Campbell-Hausdorf expansion truncates naturally after the fourfold commutator as in standard electronic CC theory. 
With a converged set of amplitudes, the CC correlation energy $\Delta E_\text{CC}$ is obtained by projecting on the reference state
\begin{equation}
	\begin{split}
      \Delta E_\text{CC} & = \bra{\text{0,0}}\text{e}^{-\hat{Q}} \hat{H}_\text{N} \text{e}^{\hat{Q}}\ket{\text{0,0}}\;. \\
\end{split}
	\label{eq:cc-energy}
\end{equation}
For the diagrammatic representation of the CC correlation energy in Eq. (\ref{eq:cc-energy}), all closed contractions of the normal-ordered  Hamiltonianian must be found, leaving no open lines, see Fig.\ref{fig:energy}. 
\\
These diagrams are evaluated as 
\begin{equation}
	\begin{split}
      \Delta E_\text{CC} & =  \sum_{ai} \tilde{f}_{ia} t_{ai} + \sum_{abij} \left( \frac{1}{4} t_{ijab} + \frac{1}{2}t_{ia}t_{jb}\right)\tilde{g}_{abij} \\
      & +  \sum_{ai}\sum_{\alpha} \left( s_{ia}^\alpha + t_{ia}\gamma^\alpha \right) \tilde{d}_{ai}^\alpha + \sum_{\alpha} \sum_{i} \tilde{d}_{ii}^\alpha \gamma^{\alpha}\;, \\
	\end{split}
	\label{eq:energy}
\end{equation}
where the first and second terms correspond to the standard electronic CC correlation energy (first three diagrams) and the third and fourth terms to the interaction with the  field (last three diagrams).
Note that the integrals for the Fock operator $\tilde{f_{ia}}$ and for the two-electron integrals $\tilde{g}_{abij}$ include the dipole self-energy.
Similar to regular CC theory, the diagrams in Fig. \ref{fig:energy} are independent of the truncation of the cluster operator $\hat{Q}$. 
This means that even when higher excitations are included in $\hat{Q}$ (like $\hat{T}_3, \hat{S}_3^1 \text{ or } \hat{\Gamma}_2$) the energy is calculated from the diagrams given in Fig. (\ref{fig:energy}).
For the interpretation of the diagrams no additional rules are needed. 
Since photonic lines always refer to quasiparticles, there is no need to adapt the sign or prefactor of a diagram due to photonic contractions.
It is also not necessary to introduce a prefactor different from one due to photonic contractions, as amplitudes with photonic indices can never be contracted in a symmetric manner. 
The bubble diagram is evaluated as 
\begin{equation}
  \begin{minipage}{0.2\linewidth}
 \centering
      \begin{fmffile}{energy-bubble}
       \begin{fmfgraph*}(25,25)
      \fmfstraight
      \fmftop{t1,t2,t3}
      \fmfbottom{b1,b2,b3}
      \fmf{phantom}{b1,m1,t1}
      \fmf{phantom}{b2,m2,t2}
      \fmf{phantom}{b3,m3,t3}
      \fmf{photon,tension=.0}{b1,m1}
      \fmf{dashes,tension=.0}{m1,m2}
      \fmffreeze
       \fmfshift{(4.6,0)}{m3}
      \fmf{fermion,left}{m3,m2,m3}
      \fmfhline{b1}
      \fmfdot{m1,m2}
       \end{fmfgraph*}
      \end{fmffile}
       \end{minipage}
		 \leftrightarrow \sum_{i\alpha} \gamma_{\alpha} d_{ii\alpha}  = \sum_{\alpha} \gamma_\alpha \braket{d}_{\alpha }\;,
	\label{}
\end{equation}
with the sum running over occupied orbitals and the modes $\alpha$.
However, as discussed before all contractions including bubble diagrams vanish for coherent-state HF orbitals.

The projected amplitude equations (\ref{eq:cc-amplitudes}) can be set up in a similar manner, where open lines represent the target indices of the excitation $\bra{\mu,\nu}$.  
Note that also the $T_1$ and $T_2$ amplitude equations have to be adapted by additional contractions over photonic operators. 
These are obtained by projecting equation (\ref{eq:cc-amplitudes}) on electronic single and double excited state with the EM-field in its vacuum $\bra{\mu,0}$.

\begin{figure}
\begin{framed}
\include{include/T1-amplitudes.tex}
\end{framed}
\caption{Additional diagrams for T1-Amplitude equation. The diagramms containing only electronic contractions are not given.}
\label{fig:T1}
\end{figure}

\begin{figure*}
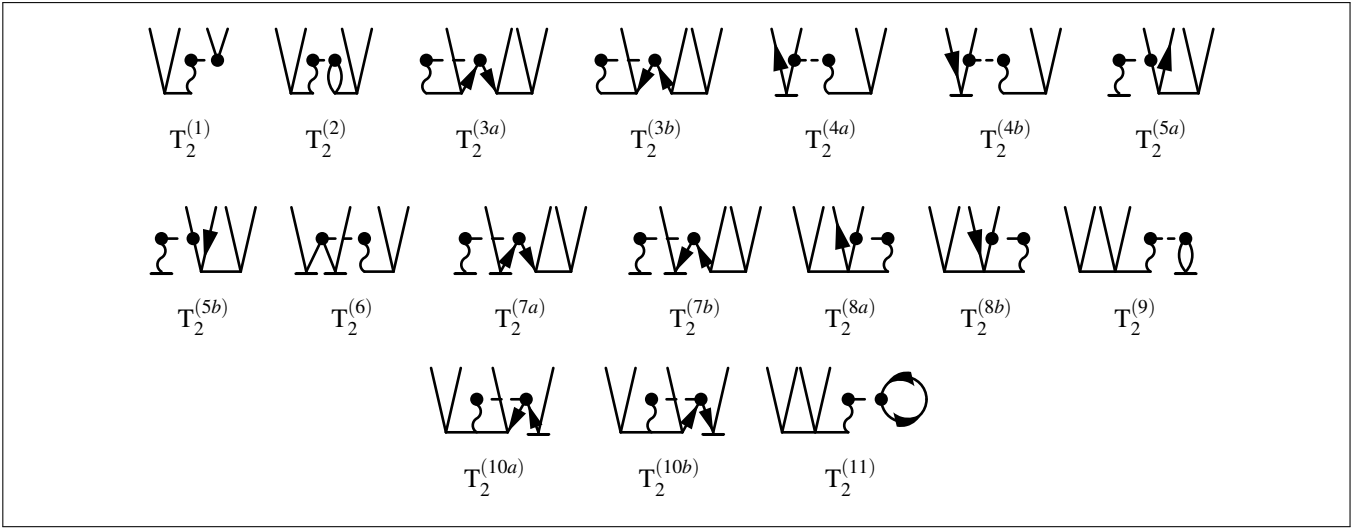

\begin{framed}
\include{include/T2-amplitudes.tex}
\end{framed}
\caption{Additional diagrams for T2-Amplitude equation. The diagrams containing only electronic contractions are not given.}
\label{fig:T2}
\end{figure*}

The prefactors in in Fig. \ref{fig:T1} are derived from the standard diagrammatic rules by analyzing the contractions for electronic indices.  
These diagrams can easily be implemented on top of existing contractions of a standard electronic CCSD code.
As an alternative, an implementation based on a $T_1$ transformed Hamiltonian can be performed.\cite{koch1994t1}
In a similar manner the additional diagrams for the $T_2$ amplitude equations are obtained which are given in Fig. \ref{fig:T2}.

In order to show how diagrams including photonic contractions are evaluated, a few contractions will be given explicitly.
E.g., the diagram $\text{T}_2^{(3a)}$ is evaluated as
\begin{equation}
     \begin{minipage}{0.23\linewidth}
      \centering
     \begin{fmffile}{T2-C1-diagram2}
      \begin{fmfgraph*}(40,25)
     \fmfstraight
     \fmftop{t1,t2,t3,t4}
     \fmfbottom{b1,b2,b3,b4}
     \fmf{phantom}{b1,m1,t1}
     \fmf{photon,tension=.0}{b1,m1}
     \fmf{phantom}{b2,t2}
     \fmfi{plain}{vpath (__b2,__t2) rotatedaround(vloc(__b2), 13)} 
      \fmffreeze
     \fmf{phantom}{b3,t3}
     \fmfi{plain}{vpath (__b3,__t3) rotatedaround(vloc(__b3),-13)} 
     \fmf{phantom}{b3,m3,t3}
     \fmffreeze
     \fmf{phantom}{b4,t4}
     \fmfi{plain}{vpath (__b4,__t4) rotatedaround(vloc(__b4),-13)} 
     \fmfi{plain}{vpath (__b4,__t4) rotatedaround(vloc(__b4),13)} 
     \fmffreeze
     \fmf{phantom}{b2,m2,t2}
     \fmffreeze
     \fmfshift{(7,0)}{m2}
     \fmf{fermion}{b2,m2}
     \fmf{fermion}{m2,b3}
     \fmfdot{m1,m2}
     \fmf{dashes}{m1,m2}
     \fmf{plain}{b1,b2}
     \fmf{plain}{b3,b4}
      \end{fmfgraph*}
     \end{fmffile}
      \end{minipage}
		\leftrightarrow -P(ij)\sum_{em\beta} s_{ei}^\beta \; t_{abmj} \; d_{me}^\beta\;
	\label{eq:T2-C1}
\end{equation}
with $P(ij)$ being the operator permuting the indices $i\leftrightarrow j$.
The prefactor of $(-1)$  derives naturally from analyzing the electronic contractions in the usual manner, where the number of hole lines is subtracted from the number of loops and the photonic contraction can be ignored.
Similarly, there is no additional prefactor for contracting the amplitude $s_{ei}^\beta$ with the Hamiltonian via ($e \text{ and } \beta$) as the two indices refer to electrons and photons respectively.

\begin{figure}
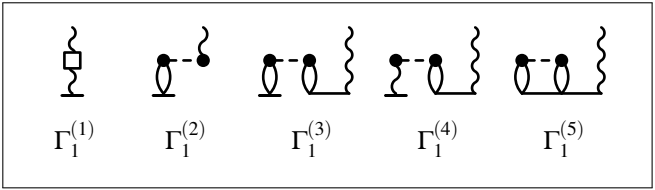

\begin{framed}
\include{include/G1-amplitudes.tex}
\end{framed}
\caption{Diagrams to determine the $\Gamma_1$-amplitudes.}
\label{fig:G1}
\end{figure}

Besides the adapted $T_1$ and $T_2$ amplitude equations, the amplitude equations for the photonic amplitudes and mixed amplitudes must be solved by projecting equation (\ref{eq:cc-amplitudes}) on $\bra{0,1}$,  $\bra{S,1}$ and $\bra{D,1}$.
The computation of the contractions for the photonic amplitudes which are given in Fig. \ref{fig:G1} is comparatively cheap as the most expensive contraction ($\Gamma_1^{(5)}$) scales with $N_{\mathrm{occ}}^2 N_{\mathrm{virt}}^2 N_{\mathrm{cav}}$ where $N_\mathrm{cav}$ corresponds to the number of modes.

\begin{figure*}
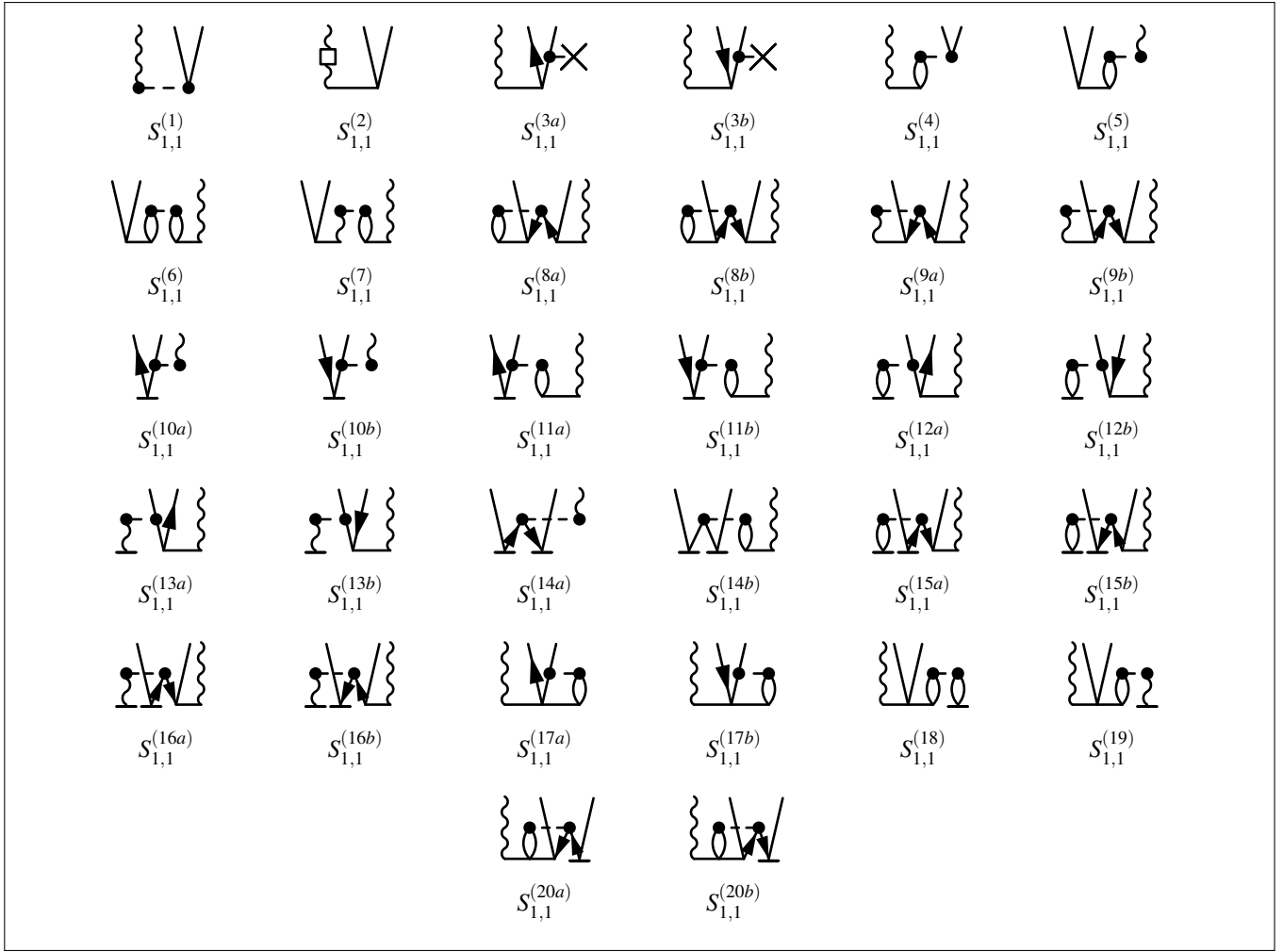

\begin{framed}
\include{include/S11-amplitudes.tex}
\end{framed}
\caption{Diagrams to determine the  $S_1^1$ amplitudes.}
\label{fig:S11}
\end{figure*}

The first diagram $\Gamma_1^\text{(1)}$ is the contraction of the $\hat{\Gamma}_1$ operator with the photonic operator $\hat{H}_\text{ph}$ and is evaluated as
\begin{equation}
     \begin{minipage}{0.05\linewidth}
    \begin{fmffile}{G1-A-diagram}
     \begin{fmfgraph*}(10,25)
     \fmfstraight
    \fmftop{t1}
    \fmfbottom{b1}
    \fmf{phantom}{b1,m1,t1}
    \fmf{photon}{b1,t1}
    \fmfhline{b01}
    \fmfv{decor.shape=square,decor.filled=empty,decor.size=3thick}{m1}
     \end{fmfgraph*}
     \end{fmffile}
     \end{minipage}
	\leftrightarrow \sum_\alpha \omega_\alpha \gamma_\alpha
	\label{eq:G1A}
 .
\end{equation}
All diagrams in FIG (\ref{eq:G1A}) have a prefactor of one, except the last diagram $\Gamma_1^\text{(5)}$ which has a prefactor of $\frac{1}{4}$ (from the symmetric contraction of electronic indices).
The contractions for the mixed amplitudes $S_1^1$ and $S_2^1$ given in Fig. \ref{fig:S11} and Figs. \ref{fig:S21-1}-\ref{fig:S21-2} are somewhat more involved.
The $S_2^1$ amplitude contractions must be implemented in an efficient manner using intermediates and factorizations, to not break the $N^6$ scaling of a cavity-free CCSD calculation. 
At this point, it should be emphasized that for QED-CCSD-1-SD the most  costly contraction if $N_\text{el} \gg N_\text{cav}$ (which is typically the case) scales as $N_\text{cav}\cdot N_\text{p}^4\cdot N_\text{h}^2$. 
In the single-mode approximation,  CCSD-1-SD  hence scales with $N^6$.
Still, a QED-CCSD-1-SD calculation involves about twice the computational cost compared to a standard CCSD calculation since the most expensive contractions with the four center Hamiltonian $\tilde{g}_{abcd}$ must be formed twice:
\begin{equation}
\begin{minipage}{0.20\textwidth}
 \centering
\begin{fmffile}{S21-4particle-diagram}
 \begin{fmfgraph*}(35,25)
\fmfstraight
\fmftop{t1,t2,t3}
\fmfbottom{b1,b2,b3}
\fmf{photon}{b1,t1}
\fmf{phantom}{b2,m2,t2}
\fmf{phantom}{t2,b2}
 \fmfi{plain}{vpath (__b2,__t2) rotatedaround(vloc(__b2),-13)}
 \fmfi{fermion}{vpath (__b2,__t2) rotatedaround(vloc(__b2),13)}
 \fmffreeze
 \fmfshift{(3,0)}{m2}
\fmf{phantom}{b3,m3,t3}
\fmf{phantom,tension=.0}{t3,b3}
 \fmffreeze
 \fmfi{plain}{vpath (__b3,__t3) rotatedaround(vloc(__b3), 13)}
 \fmfi{fermion}{vpath (__b3,__t3) rotatedaround(vloc(__b3),-13)}
 \fmffreeze
 \fmfshift{(-3,0)}{m3}
\fmf{plain}{b1,b3}
\fmf{dashes,tension=.0}{m2,m3}
 \fmfdot{m2,m3}
 \end{fmfgraph*}
\end{fmffile}
 \end{minipage}
 \text{\&}
 \begin{minipage}{0.20\textwidth}
 \centering
\begin{fmffile}{T2-4particle-diagram}
 \begin{fmfgraph*}(22,25)
\fmfstraight
\fmftop{t2,t3}
\fmfbottom{b2,b3}
\fmf{phantom}{b2,m2,t2}
\fmf{phantom}{t2,b2}
 \fmfi{plain}{vpath (__b2,__t2) rotatedaround(vloc(__b2),-13)}
 \fmfi{fermion}{vpath (__b2,__t2) rotatedaround(vloc(__b2),13)}
 \fmffreeze
 \fmfshift{(3,0)}{m2}
\fmf{phantom}{b3,m3,t3}
\fmf{phantom,tension=.0}{t3,b3}
 \fmffreeze
 \fmfi{plain}{vpath (__b3,__t3) rotatedaround(vloc(__b3), 13)}
 \fmfi{fermion}{vpath (__b3,__t3) rotatedaround(vloc(__b3),-13)}
 \fmffreeze
 \fmfshift{(-3,0)}{m3}
\fmf{plain}{b2,b3}
\fmf{dashes,tension=.0}{m2,m3}
 \fmfdot{m2,m3}
 \end{fmfgraph*}
\end{fmffile}
 \end{minipage} .
 \end{equation}
The memory requirement on the other hand is comparable to standard CCSD when efficiently organizing the intermediates and factorizations and when $N_\text{el} \gg N_\text{cav}$.

As usual, the amplitude equations are solved in an iterative manner by subtracting the contraction with lowest order from equation (\ref{eq:cc-amplitudes}) and dividing by the integrals.
For the $T_1$ and $T_2$ amplitude equation the only modification is the use of the dipole self-energy terms in the Fock matrix. 
The $\Gamma_1$ amplitude equations are solved via
\begin{equation}
   \gamma^\alpha  =  \frac{\Gamma_1 \; \text{contractions}}{-\omega_\alpha}.
   \label{}
\end{equation}
In the numerator all contractions from Fig. \ref{fig:G1} are contained except for the $\Gamma_1^\text{A}$ diagramm. 
For mixed amplitudes 
we obtain
\begin{equation}
   s^\alpha_{ia}  =  \frac{S_{1,1}\;  \text{contractions}}{\tilde{f}_{ii} - \tilde{f}_{aa} - \omega_\alpha}
   \label{},
\end{equation}
\begin{equation}
   s^\alpha_{ijab}  =  \frac{S_{2,1}\;  \text{contractions}}{\tilde{f}_{ii} + \tilde{f}_{jj} - \tilde{f}_{aa} - \tilde{f}_{bb}- \omega_\alpha}
   \label{}
   ,
\end{equation}
where the numerator contains all diagrams from Fig. \ref{fig:S11} and Figs. \ref{fig:S21-1}-\ref{fig:S21-2} except for $S_{1,1}^\text{(2)}$, $S_{1,1}^\text{(3a)}$, and $S_{1,1}^\text{(3b)}$ for the $S_{1,1}$ amplitude equations  and $S_{2,1}^\text{(3)}$, $S_{2,1}^\text{(3a)}$ and $S_{2,1}^\text{(3b)}$ for the $S_{2,1}$ amplitude equations.
Note that for canonical orbitals $\tilde{f}_{ii}$ and $\tilde{f}_{aa}$ are the HF orbital energies $\epsilon_i$ and $\epsilon_a$.

\subsection{$\Gamma_2$  amplitude equations}
\label{sec:Gamma2}
\begin{figure}
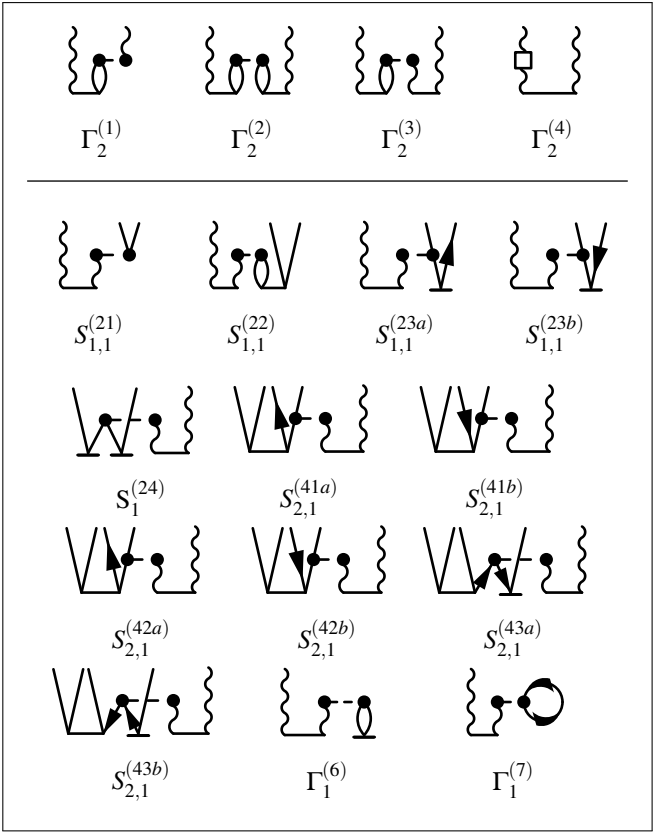

\begin{framed}
\include{include/G2_amplitudes.tex}
\end{framed}
\caption{Diagrams to determine the  $\Gamma_2$ amplitudes are shown in the first row. Diagrams that contribute to $S_{1,1}$, $S_{2,1}$ and $\Gamma_1$ equations are shown in the lower part.}
\label{fig:G2}
\end{figure}

Philbin et al. showed that including two photonic excitations in the cluster operator ($\Gamma_2$ amplitudes) is crucial for the correct description of the interaction energy in the dissociation limit.\cite{philbin2023molecular}
The corresponding truncation scheme for the cluster operator $\hat{Q}$ is called the CCSD-12-SD truncation and can easily be implemented on top an CCSD-1-SD implementation without increasing the overall scaling.
The $\Gamma_2$ operator is defined as
\begin{equation}
   \hat{\Gamma}_2 = \sum_{\alpha<\beta} \gamma^{\alpha\beta} \hat{\alpha}^\dagger \hat{\beta}^\dagger\;. 
\end{equation}
The diagrammatic rules assume antisymmetric amplitudes for which the restricted sums can be recast into unrestricted ones by the simple use of a factor of  $\frac{1}{2}$   due to the fact that the diagonal vanishes anyway ($\sum_{p >q} \rightarrow \frac{1}{2}\sum_{pq}$). Since this is no longer the case for photonic indices, we replace $\gamma^{\alpha\beta}$ with  
\begin{equation}
	\tilde{\gamma}^{\alpha\beta} = (1 + \delta_{\alpha\beta})  \gamma^{\alpha\beta}\;
	\label{eq:rescale_G2}
\end{equation}
such that the unrestricted sum can be employed 
\begin{equation}
   \hat{\Gamma}_2 = \frac{1}{2}\sum_{\alpha\beta} \tilde{\gamma}^{\alpha\beta} \hat{\alpha}^\dagger \hat{\beta}^\dagger \leftrightarrow 
       \begin{minipage}{0.20\linewidth}
       \centering
       \begin{fmffile}{G2-operator}
        \begin{fmfgraph*}(15,30)
        \fmfstraight
		  \fmftop{t1,t2}
		  \fmfbottom{b1,b2}
			 \fmf{plain}{b1,b2}
			 \fmf{photon}{b2,t2}
			 \fmf{photon}{b1,t1}
        \end{fmfgraph*}
        \end{fmffile}
	  \end{minipage}\;\\.
\end{equation}
No additional rules are then needed for the evaluation of contractions including $\Gamma_2$. 
This scheme is readily generalized to cluster operators of higher excitation level like $\Gamma_3$ or mixed cluster operators with two or more photonic indices as for example $S_1^2$.\\
The $\Gamma_2$ operator does not contribute to the CC-energy as it cannot be fully contracted with the Hamiltonian.
The corresponding diagrams to determine the $\Gamma_2$ amplitudes are given in the first row in Fig.  (\ref{fig:G2}) and additional contractions which have to be included in the $S_{1,1}$, $S_{2,1}$, and $\Gamma_1$ equations are given below.
Also here, all contractions use $\tilde{\gamma}_{\alpha\beta}$ instead of $\gamma_{\alpha\beta}$ amplitudes as discussed before.
As an example, the $\Gamma_2^Z$ diagram is evaluated as
\begin{equation}
\begin{minipage}{0.25\linewidth}
 \centering
\begin{fmffile}{G2-test-diagram}
 \begin{fmfgraph*}(30,25)
\fmfstraight
\fmftop{t1,t2,t3,t4}
\fmfbottom{b1,b2,b3,b4}
\fmf{phantom}{b2,m2,t2}
\fmf{vanilla,left=.4,tension=.0}{b2,m2}
\fmf{vanilla,left=.4,tension=.0}{m2,b2}
\fmf{phantom}{b3,m3,t3}
\fmf{photon,tension=.0}{b3,m3}
\fmf{dashes,tension=.0}{m2,m3}
 \fmfdot{m2,m3}
 \fmf{plain}{b1,b2}
 \fmf{plain}{b3,b4}
 \fmf{photon}{b4,t4}
\fmffreeze
\fmf{photon}{b1,t1}
 \end{fmfgraph*}
\end{fmffile}
 \end{minipage}
  \leftrightarrow \mathcal{P}(\alpha\beta)\sum_{ai}\sum_{\lambda} s_{ai}^\alpha \tilde{\gamma}^{\lambda\beta} d_{ai}^\lambda 
\end{equation}
where the permutation generated by $\mathcal{P}(\alpha\beta)$ is always symmetric for bosonic indices.
The symmetric permutation also takes care of scaling the diagonal correctly since for the same index we have $\mathcal{P}(\alpha\alpha) = 2$.
The most expensive contraction for the $\Gamma_2$ amplitudes scales with $N^5$ and does hence not increase the scaling of CCSD-1-SD. 

\section{QED equation-of-motion coupled cluster}
\label{sec:eom}
\begin{figure}
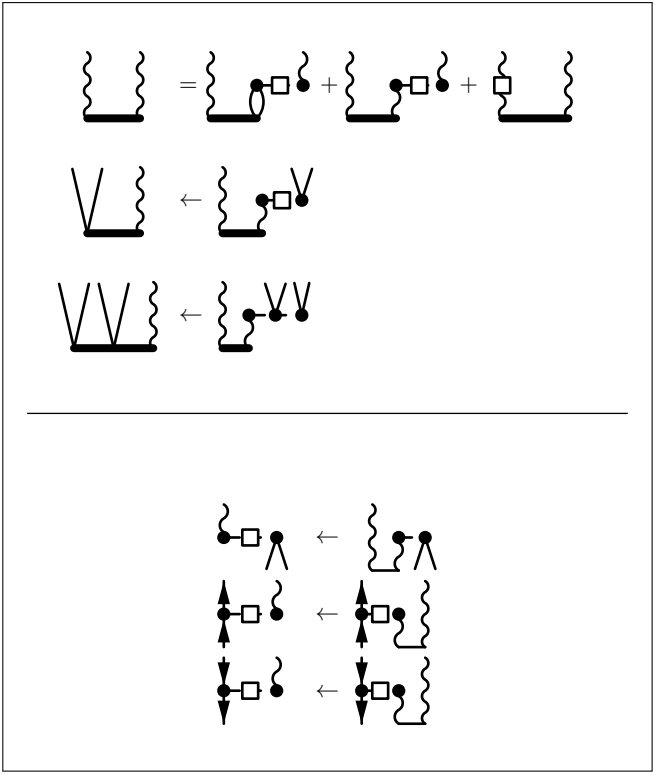

\begin{framed}
\include{include/eom-ccsd-12-sd.tex}
\end{framed}
\caption{Top - Equations for the EOM-CCSD-12-SD truncation; Bottom - Additional intermediates. }
\label{fig:eom-ccsd-12-sd}
\end{figure}

In cavity-QED electronically excited states play an important role as they are even more susceptible to the light matter coupling than ground states.\cite{ruggenthaler2023understanding}
Starting from a QED-CC reference wave function, excited states can be generated using a CI-like parametrization 
\begin{equation}
   \hat{\tilde{H}}_{\text{N}} \hat{R} \ket{0,0} = \Delta E_\text{exc}\hat{R} \ket{0,0}\;
   \label{eq:EOM}
\end{equation}
where the similarity-transformed Hamiltonian in normal order was introduced
\begin{equation}
   \hat{\tilde{H}}_{\text{N}} = \text{e}^{-\hat{Q}} \hat{H}_\text{N} \text{e}^{\hat{Q}}\;.
   \label{eq:hbar}
\end{equation}
The excitation operator $\hat{R}$ is truncated in the same manner as the cluster operator $\hat{Q}$.
For the CCSD-1-SD truncation scheme it is therefore given as
\begin{equation}
   \hat{R} = r_{0,0} + \hat{R}_{1,0} + \hat{R}_{2,0} + \hat{R}_{1,1} + \hat{R}_{2,1} + \hat{R}_{0,1}\;,
   \label{}
\end{equation}
where the first index refers to the electronic excitation level and the second to the number of generated photons.
The mixed and photonic excitation operators are written in second quantization and diagrammatically as 
\begin{align}
   \hat{R}_{1,1} &= \sum_{ai\alpha} r_{ai}^\alpha \hat{a}^\dagger \hat{i}\hat{\alpha}^\dagger  &\leftrightarrow
       \begin{minipage}{0.30\linewidth}
       \centering
       \begin{fmffile}{RS1-operator}
        \begin{fmfgraph*}(15,30)
        \fmfstraight
		  \fmftop{t1,t2}
		  \fmfbottom{b1,b2}
		    \fmf{phantom}{b1,t1}
			 \fmfi{fermion}{vpath (__b1,__t1) rotatedaround(vloc(__b1),-13)}
		    \fmf{phantom}{t1,b1}
			 \fmfi{fermion}{vpath (__t1,__b1) rotatedaround(vloc(__b1),13)}
          \fmf{plain,width=3}{b1,b2}
			 \fmf{photon}{b2,t2}
        \end{fmfgraph*}
        \end{fmffile}
	  \end{minipage}\;,\\
     \hat{R}_{2,1} &= \frac{1}{4} \sum_{abij\alpha} r_{abij}^\alpha \hat{a}^\dagger\hat{b}^\dagger \hat{j}\hat{i} \hat{\alpha}^\dagger &\leftrightarrow 
       \begin{minipage}{0.30\linewidth}
       \centering
       \begin{fmffile}{RS2-operator}
        \begin{fmfgraph*}(30,30)
        \fmfstraight
		  \fmftop{t1,t2,t3}
		  \fmfbottom{b1,b2,b3}
		    \fmf{phantom}{b1,t1}
			 \fmfi{fermion}{vpath (__b1,__t1) rotatedaround(vloc(__b1),-13)}
		    \fmf{phantom}{t1,b1}
			 \fmfi{fermion}{vpath (__t1,__b1) rotatedaround(vloc(__b1),13)}
		    \fmf{phantom}{b2,t2}
			 \fmfi{fermion}{vpath (__b2,__t2) rotatedaround(vloc(__b2),-13)}
		    \fmf{phantom}{t2,b2}
			 \fmfi{fermion}{vpath (__t2,__b2) rotatedaround(vloc(__b2),13)}
          \fmf{plain,width=3}{b1,b3}
			 \fmf{photon}{b3,t3}
        \end{fmfgraph*}
        \end{fmffile}
	  \end{minipage}\;,\\
     \hat{R}_{0,1} &= \sum_{\alpha} r^\alpha \hat{\alpha}^\dagger &\leftrightarrow
       \begin{minipage}{0.30\linewidth}
       \centering
       \begin{fmffile}{RG1-operator}
        \begin{fmfgraph*}(10,30)
        \fmfstraight
		  \fmftop{t1,t2,t3}
		  \fmfbottom{b1,b2,b3}
		    \fmf{photon}{b1,t1}
			 \fmfdhline{b1}
        \end{fmfgraph*}
        \end{fmffile}
	  \end{minipage}\;.
	  \label{eq:cluster-op}
\end{align}
The bold line is used to distinguish the $\hat{R}$ operator from the cluster operator $\hat{Q}$.
The diagonalization of the full similarity-transformed Hamiltonian matrix is usually avoided by only calculating the lowest eigenvalues in a Davidson-like procedure. 
When further exploiting the block diagonal structure of the similarity transformed Hamiltonian, states of different symmetry can be calculated separately.
The QED-EOM-CCSD-1-SD diagrams are given in Fig. \ref{fig:EOM} and the corresponding intermediates are found in Figs. \ref{fig:Int1}-\ref{fig:Int3}.
The additional diagrams needed for a QED-EOM-CCSD-12-SD calculation are given in Fig. \ref{fig:eom-ccsd-12-sd}. 

\begin{figure*}
\begin{framed}
\include{include/eom.tex}
\end{framed}
\caption{EOM-CCSD-1-SD equation}
\label{fig:EOM}
\end{figure*}

\begin{figure*}
\begin{framed}
\include{include/intermediates1.tex}
\end{framed}
\caption{Intermediates for the EOM-CCSD-1-SD truncation.}
\label{fig:Int1}
\end{figure*}
\begin{figure*}
\begin{framed}
\include{include/intermediates2.tex}
\end{framed}
\caption{Intermediates for the EOM-CCSD-1-SD truncation.}
\label{fig:Int2}
\end{figure*}
\begin{figure*}
\begin{framed}
\include{include/intermediates3.tex}
\end{framed}
\caption{Intermediates for the EOM-CCSD-1-SD truncation.}
\label{fig:Int3}
\end{figure*}

\subsection{CI-type initial guess for the EOM vectors}
In photon-free EOM-CC implementations the eigenvectors of the configuration interaction with singles (CIS) matrix are often used as initial guess for the coefficients $r_{1,0}$ and $r_{2,0}$ in $\hat{R}$.
This enables the targeted calculation of selected electronic states, especially when also making use of point-group symmetry. 
For QED-EOM-CC additional projections onto photonic excited states are required.
Here, we discuss the single-mode approximation including hence only one photonic index. However, the expressions are readily generalized to more cavity modes.

To obtain a guess for a QED-EOM-CC treatment, the CIS matrix is expanded by the $\ket{0,1}$ state (containing a single photon) resulting in 
\begin{equation}
   \begin{split}
   \begin{pmatrix}
      \braket{0,0|H_N|0,0}  & \braket{0,0|H_N|S,0} & \braket{0,0|H_N|0,1}\\
      \braket{S,0|H_N|0,0}  & \braket{S,0|H_N|S,0} & \braket{S,0|H_N|0,1}\\
      \braket{0,1|H_N|0,0}  & \braket{0,1|H_N|S,0} & \braket{0,1|H_N|0,1}\\
   \end{pmatrix}\\
   =
   \begin{pmatrix}
      E_\text{HF} & 0 & 0 \\
      0           & \boldsymbol{H}_\text{S}  & \boldsymbol{d}\\
      0           & \boldsymbol{d}^\dagger & E_\text{HF} + \omega\\
   \end{pmatrix}
   \end{split}
   \label{eq:tg_cis_mat}
\end{equation}
where the off-diagonal blocks in the electronic part vanish due to Brillouin's theorem.
The electronic states $\ket{S,0}$ and the photonic state $\ket{0,1}$ are coupled via the transition dipole moments
\begin{equation}
   \begin{split}
      \tilde{d}_{ia} &= \braket{i\rightarrow a,0|H|0,1} \;.
   \end{split}
   \label{}
\end{equation}
Note that the off-diagonal elements $\braket{0,0|H|0,1}$ vanish in the coherent-state basis
\begin{equation}
   \braket{0,0|H|0,1} = \braket{\hat{\tilde{d}}} = \braket{d-\braket{d}} = 0
   \label{}
\end{equation}
which can be seen as a generalization of Brillouin's theorem for the photonic degree of freedom.
The diagonalization of the extended CIS matrix in Eq.  (\ref{eq:tg_cis_mat}) yields an additional element in the eigenvector $\tilde{\boldsymbol{r}}_i$ which is used as guess for the photonic coefficient $r^1$.
These eigenvectors yield suitable guess vectors for the upper and lower polariton in the system which are mainly linear combinations of electronically excited states $\ket{S,0}$ and the photonic state $\ket{0,1}$. 
However, they do not describe states where both, an electronic state is excited together with a photon creation (e.g. $\ket{S,1}$) as this corresponds to a double excitation. 
In order to also be able to obtain a guess for such states as well, the CIS matrix is further expanded by states of the type $\ket{S,1}$ into a QED-CIS(D)-type matrix
\begin{equation}
   \begin{split}
   &\begin{pmatrix}
      \braket{0,0|H|0,0}  & \braket{0,0|H|S,0} & \braket{0,0|H|0,1} & \braket{0,0|H|S,1}\\
      \braket{S,0|H|0,0}  & \braket{S,0|H|S,0} & \braket{S,0|H|0,1} & \braket{S,0|H|S,1}\\
      \braket{0,1|H|0,0}  & \braket{0,1|H|S,0} & \braket{0,1|H|0,1} & \braket{0,1|H|S,1}\\
      \braket{S,1|H|0,0}  & \braket{S,1|H|S,0} & \braket{S,1|H|0,1} & \braket{S,1|H|S,1}\\
   \end{pmatrix}\\
   &=
   \begin{pmatrix}
      E_\text{HF} & 0 & 0 & \boldsymbol{d}^\dagger \\
      0  & \boldsymbol{H}_\text{S}  & \boldsymbol{d} & \boldsymbol{D}  \\
      0  & \boldsymbol{d}^\dagger & E_\text{HF} + \omega & 0      \\
      \boldsymbol{d}  & \boldsymbol{D}^\dagger & 0 &\boldsymbol{H}_\text{S} + \mathbb{1} \omega       \\
   \end{pmatrix}.
   \end{split}
   \label{eq:tgs_cis_mat}
\end{equation}
The elements $\braket{0,1|H|S,1}$ vanish due to the Brillouin theorem.
The four-index tensor $D_{ijab}$ contains the transition-dipole moments between different electronically excited states
\begin{align}
   (\boldsymbol{D})_{ijaa} &= \braket{i\rightarrow a|\hat{\tilde{d}}|j\rightarrow a} = \tilde{d}_{ij}\\
   (\boldsymbol{D})_{iiab} &= \braket{i\rightarrow a|\hat{\tilde{d}}|i\rightarrow b} = \tilde{d}_{ab}\\
   (\boldsymbol{D})_{ijab} &= 0 \qquad \text{for} \quad i\ne j \quad \text{and} \quad a\ne b 
   \label{}
\end{align}
From the matrix (\ref{eq:tgs_cis_mat}) it becomes obvious that the inclusion of $\ket{S,1}$ causes couplings to the reference state $\ket{0,0}$. 
The eigenvectors of the Hamiltonian in Eq. (\ref{eq:tgs_cis_mat}) are therefore only considered suitable as initial guess for $\hat{R}$ as long as couplings to the ground state are small. 
If also $\Gamma_2$ amplitudes are included, the CIS guess can further be generalized by also projecting onto the doubly excited photonic states  $\ket{0,2}$, yielding 
\begin{equation}
   \begin{split}
   \boldsymbol{H}_{CIS(D)} = 
   \begin{pmatrix}
      E_\text{HF}                  &  0                    & 0 & 0 & \boldsymbol{d}^\dagger \\
      0  & \boldsymbol{H}_\text{S} & \boldsymbol{d}        & 0 & \boldsymbol{D}  \\
      0  & \boldsymbol{d}^\dagger  & E_\text{HF} + \omega    & 0 & 0      \\
      0  & 0                       & 0                       & E_\text{HF} + 2 \omega   & \boldsymbol{d}^\dagger \\
      \boldsymbol{d}               & \boldsymbol{D}^\dagger & 0 &  \boldsymbol{d}  &\boldsymbol{H}_\text{S} + \mathbb{1} \omega &     \\
   \end{pmatrix}
   \end{split}
   \label{eq:tgs_cis_mat2}
   .
\end{equation}
This scheme can further be generalized to more photons.

When employing point-group symmetry, the $\boldsymbol{H}_\text{CIS(D)}$ matrix can be diagonalized for individual symmetry blocks separately.
Note that in this case the two submatrices $\boldsymbol{H}_\text{S}$ appearing in $\boldsymbol{H}_\text{CIS(D)}$ do not have to belong to the same symmetry blocks. 

\section{Point-group symmetry in Cavities}
\label{sec:symmetry}
Exploiting point-group symmetry in the context of direct-product decomposition in CC theory leads to significant speedups in computational time\cite{stanton1991symmetry} and enables the targeted calculation of states of different irreducible representations.
Here we discuss the symmetry of molecules in cavities and the corresponding approximations in quantum-chemical calculations.

\subsection{The isolated electromagentic field}
\begin{figure}
    \centering
    \includegraphics[width=0.6\linewidth]{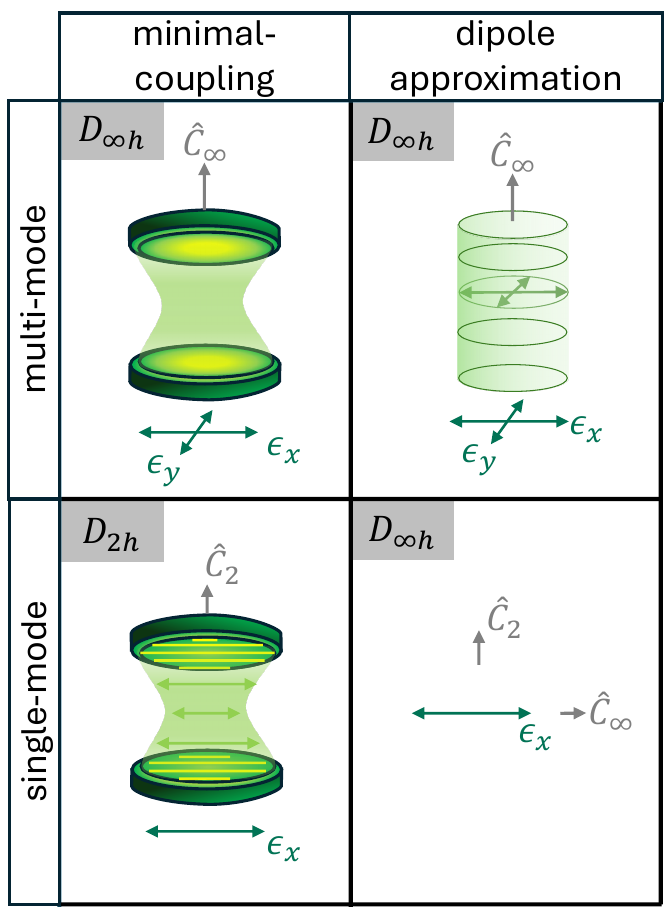}
    \caption{Symmetry of the cavity in the minimal coupling picture vs. the dipole approximation and multi-mode vs. single-mode approximation.}
    \label{fig:cavity-sym}
\end{figure}

Depending on the approximation in the Hamiltonian, the cavity symmetry is different as shown in Fig. \ref{fig:cavity-sym}. 
In the upper left corner, a general one-dimensional cavity of arbitrary polarization is depicted which is characterized by the principle rotation axis $C_\infty$  pointing in the direction of the two mirrors. The  system has an overall $D_{\infty h}$ symmetry. 
Obviously, when the EM field is polarized, the symmetry is reduced as for example for linearly  polarized light with polarization direction $\boldsymbol{\epsilon}_x$ - see lower left panel in Fig. \ref{fig:cavity-sym}. In a quantum-chemical calculation the latter situation can be described within the single-mode approximation. 
Note that the term single-mode approximation is somewhat ambiguous. In the physical sense what is meant is that  the EM field is described only by its fundamental frequency.  This approximation is motivated by the fact that all other modes are energetically higher and do not lead to significant interactions. However, the EM field in its fundamental frequency is still composed of two orthogonal modes of different field polarization. 
The principle rotation axis is then reduced from $C_\infty$ to $C_2$, noting that two orthogonal $C_2$ rotations remain.
Experimentally both, rotationally symmetric cavities as well as cavities with predefined field polarization can be realized. 
When applying the dipole approximation it is assumed that the wavelength of the EM field is large over the dimension of the molecule, meaning that in the limit, the two mirrors are infinitely far apart. In Fig. \ref{fig:cavity-sym} this is realized by removing the mirrors so that the electric field has no boundaries. 
This can be understood as an rotationally symmetric electric field pointing in an arbitrary direction in the $xy$-plane - see upper right corner of Fig. \ref{fig:cavity-sym}. 
When applying single-mode treatment within the dipole approximation, one of the polarization vectors is removed by which the electric field is given a predefined polarization. 
This can be seen in Fig. \ref{fig:cavity-sym} in the lower-right corner, where a homogeneous electric field is oriented along the $\epsilon_x$-vector.  
In consequence, an additional $C_\infty$ rotation axis is formed in the direction of the polarization vector. This means that symmetry of the system in the dipole approximation together with the single-mode treatment is higher ($D_{\infty h}$) than in the experimental setup for a linear polarized cavity ($D_{2h}$) which may lead to inaccuracies.

\subsection{The Hamiltonian in the dipole approximation}
In the dipole approximation, the terms responsible for a symmetry reduction are all operators containing photonic creation and annihilation operators ($\hat{H}_\text{bil}$ and $\hat{H}_\text{ph}$).  
The operator $\hat{H}_\text{bil}$ consist of two individual terms, each consisting of three elementary operators  ($\hat{\alpha} \hat{p}^\dagger \hat{q}$ and $\hat{\alpha}^\dagger \hat{p}^\dagger \hat{q}$).
While the electronic creation and annihilation operators transform according to the symmetry of molecular orbitals, the photonic creation and annihilation operators transform according to a homogeneous electric field pointing in the direction of $\boldsymbol{\lambda}_\alpha$. 
The EM field in the dipole approximation is assumed to be constant and thus the electric field becomes homogeneous. 
A photon created by $\hat{\alpha}^\dagger$ therefore corresponds to a homogeneous electric field aligned along $\boldsymbol{\lambda}_\alpha$.
E.g. a reflection perpendicular to the polarization vector $\boldsymbol{\lambda}_\alpha$ changes the sign
\begin{equation}
   \hat{\sigma}_h^\dagger \hat{\alpha} \hat{\sigma}_h = - \hat{\alpha}
   \label{}
\end{equation}
while a reflection with respect to a plane containing the polarization vector is symmetric
\begin{equation}
   \hat{\sigma}_v^\dagger \hat{\alpha} \hat{\sigma}_v =  \hat{\alpha}.
   \label{}
\end{equation}
Similarly, photonic operators are symmetric for an arbitrary rotation around an axis containing the polarization vector $\boldsymbol{\lambda}_\alpha$ and anti-symmetric for $C_2$ rotations perpendicular to $\boldsymbol{\lambda}_\alpha$. 

The photonic Hamiltonian $\hat{H}_\text{ph}$ always transforms in a totally symmetric manner under the symmetry operations within the group as the direct product $\Gamma_\alpha \otimes \Gamma_\alpha$ always contains the totally symmetric representation. 
For the bilinear coupling operator (\ref{eq:bil}) it holds that 
\begin{equation}
   \Gamma_1 \in \Gamma_p \otimes \Gamma_q \otimes \Gamma_\alpha, 
   \label{eq:bilirrep}
\end{equation}
i.e, the irreducible representations of $p$ and $q$ can be different. 
This means that even when the dipole operator appears explicitly in the Hamiltonian, the molecule can be totally symmetric with respect to horizontal reflections generated by $\hat{\sigma}_h$.
By this, molecules can be categorized into two groups. 
If the molecule has a dipole moment oriented along the polarization vector, the photonic index is totally symmetric.
For these molecules the bilinear coupling integrals $\tilde{d}^\alpha_{pq}$ become block diagonal in the electronic indices ($\Gamma_p = \Gamma_q$).
On the other hand, if the molecule has no dipole moment, or a dipole moment not aligned with the polarization vector, it holds for the photonic index that $\Gamma_\alpha \ne \Gamma_1$. Hence the blocks for $\Gamma_p = \Gamma_q$ vanish.
Thus the implementation needs to be flexible enough to handle both, block diagonal matrices as well as matrices with non-zero blocks on the off-diagonals. 

\subsection{Symmetry in CC theory}
Handling of symmetry based on the direct product decomposition in the context of CC theory is well known \cite{stanton1991symmetry} and has been exploited in many quantum-chemical program packages. 
It ensures that the number of floating point operations can be reduced by a factor of $h^2$, with $h$ being the order of the group. 
For use within QED-CC theory, the photonic indices must be taken into account. 

As the cluster operator $\hat{Q}$ must be totally symmetric, it is obvious that the $\gamma^\alpha$ amplitude vanishes when $\hat{\alpha}^\dagger$ is not totally symmetric itself.
In other words, the molecule must have a dipole moment along $\boldsymbol{\lambda}_\alpha$ in order to have non-vanishing contributions in $\gamma^\alpha$ (when $\alpha$ is totally symmetric the electronic indices of $\tilde{d}_{pq}^\alpha$ must also be totally symmetric which in turn is only the case if the molecule has a dipole moment oriented along $\boldsymbol{\lambda}_\alpha$).

For the same reason in the mixed amplitudes  $s_{ia}^\alpha$ and $s_{ijab}^\alpha$ the electronic indices do not have to contain the totally symmetric irreducible representation.
For the contractions it follows for example 
for Eq. (\ref{eq:T2-C1})   
\begin{equation}
     \begin{minipage}{0.23\linewidth}
      \centering
     \begin{fmffile}{T2-C1-diagram2}
      \begin{fmfgraph*}(40,25)
     \fmfstraight
     \fmftop{t1,t2,t3,t4}
     \fmfbottom{b1,b2,b3,b4}
     \fmf{phantom}{b1,m1,t1}
     \fmf{photon,tension=.0}{b1,m1}
     \fmf{phantom}{b2,t2}
     \fmfi{plain}{vpath (__b2,__t2) rotatedaround(vloc(__b2), 13)} 
      \fmffreeze
     \fmf{phantom}{b3,t3}
     \fmfi{plain}{vpath (__b3,__t3) rotatedaround(vloc(__b3),-13)} 
     \fmf{phantom}{b3,m3,t3}
     \fmffreeze
     \fmf{phantom}{b4,t4}
     \fmfi{plain}{vpath (__b4,__t4) rotatedaround(vloc(__b4),-13)} 
     \fmfi{plain}{vpath (__b4,__t4) rotatedaround(vloc(__b4),13)} 
     \fmffreeze
     \fmf{phantom}{b2,m2,t2}
     \fmffreeze
     \fmfshift{(7,0)}{m2}
     \fmf{fermion}{b2,m2}
     \fmf{fermion}{m2,b3}
     \fmfdot{m1,m2}
     \fmf{dashes}{m1,m2}
     \fmf{plain}{b1,b2}
     \fmf{plain}{b3,b4}
      \end{fmfgraph*}
     \end{fmffile}
      \end{minipage}
		=-P(ij)\sum_{\Gamma_A \Gamma_B \Gamma_C} \sum_{\substack{e \in \Gamma_A \\ m \in \Gamma_B}} \sum_{\beta \in \Gamma_C} s_{ei}^\beta \; t_{abmj} \; d_{me}^\beta\;
\end{equation}
with $\Gamma_1 \in \Gamma_A \otimes \Gamma_B \otimes \Gamma_C$ which simplifies if the molecule has a dipole moment along $\boldsymbol{\lambda}_\alpha$ to 
\begin{equation}
     \begin{minipage}{0.23\linewidth}
      \centering
     \begin{fmffile}{T2-C1-diagram2}
      \begin{fmfgraph*}(40,25)
     \fmfstraight
     \fmftop{t1,t2,t3,t4}
     \fmfbottom{b1,b2,b3,b4}
     \fmf{phantom}{b1,m1,t1}
     \fmf{photon,tension=.0}{b1,m1}
     \fmf{phantom}{b2,t2}
     \fmfi{plain}{vpath (__b2,__t2) rotatedaround(vloc(__b2), 13)} 
      \fmffreeze
     \fmf{phantom}{b3,t3}
     \fmfi{plain}{vpath (__b3,__t3) rotatedaround(vloc(__b3),-13)} 
     \fmf{phantom}{b3,m3,t3}
     \fmffreeze
     \fmf{phantom}{b4,t4}
     \fmfi{plain}{vpath (__b4,__t4) rotatedaround(vloc(__b4),-13)} 
     \fmfi{plain}{vpath (__b4,__t4) rotatedaround(vloc(__b4),13)} 
     \fmffreeze
     \fmf{phantom}{b2,m2,t2}
     \fmffreeze
     \fmfshift{(7,0)}{m2}
     \fmf{fermion}{b2,m2}
     \fmf{fermion}{m2,b3}
     \fmfdot{m1,m2}
     \fmf{dashes}{m1,m2}
     \fmf{plain}{b1,b2}
     \fmf{plain}{b3,b4}
      \end{fmfgraph*}
     \end{fmffile}
      \end{minipage}
		=-P(ij)\sum_{\Gamma_A} \sum_{\substack{e \in \Gamma_A \\ m \in \Gamma_A}} \sum_{\beta \in \Gamma_1} s_{ei}^\beta \; t_{abmj} \; d_{me}^\beta \;
\end{equation}
where the sum over $\Gamma_I$ is over all irreducible representations in the group. 
This allows to make use of the usual integral-packing strategies for the permutations of indices of the  different irreducible representations.
Similar to regular CC theory, if  $N_\mathrm{el} \gg N_\mathrm{cav}$, the most significant amount of memory can hence be saved by packing the two electron integrals $\tilde{g}_{abcd}$ and additionally the four and five index amplitudes $t_{ijab}$ and $s^\alpha_{ijab}$. Overall the memory requirements are then similar as compared to standard CC.  

\subsection{Symmetry for excited states} \label{sec:symexc}
Strong light-matter coupling and corresponding splitting can 
only occur when the isolated electronic excited state $\ket{\Gamma_B,0}$ and the cavity mode $\ket{\Gamma_A,1}$ belong to the same irreducible representation.
Here, $\Gamma_A$ is the irreducible representation of the electronic ground state and $\Gamma_B$ the irreducible representation of an electronic excitation while $\Gamma_\text{ph}$ is the irreducible representation of the cavity. 
Only if $ \Gamma_B \in \Gamma_A \otimes \Gamma_\text{ph}$ the coupling is allowed and an upper and lower polariton $\ket{P_+}$ and $\ket{P_-}$ can be formed. 
In the simplest case, these upper and lower polaritons can be approximated by a linear combination of the uncoupled states
\begin{equation}
   \begin{split}
   \ket{P_+} & = c_1 \ket{\Gamma_B,0} + c_2 \ket{\Gamma_A,1}\;,\\
   \ket{P_-} & = c_3 \ket{\Gamma_B,0} - c_4 \ket{\Gamma_A,1}\;.
   \end{split}
   \label{eq:polariton}
\end{equation}
Of course, if more states of the same irreducible representation are in the direct vicinity of the cavity frequency more states have to be included in Eq. (\ref{eq:polariton}).
The coefficients $c_i$ are heavily influenced by the frequency and coupling strength of the cavity and the 
  formation of the upper and lower polariton can be viewed as the result of an avoided crossing.  

\begin{figure}
\begin{minipage}{1.0\linewidth}
\begin{tikzpicture}[]

   \def\shiftA{1.5}
   \def\shiftB{3.1}
   \def\shiftC{4.9}

   \def\width{0.3}

   \def\ElC{3.0}
   \def\ElB{2.4}
   \def\ElA{0.2}

   \def\PoD{3.4}
   \def\PoC{3.1}
   \def\PoB{1.8}
   \def\PoA{0.15}

   \def\PhA{2.8}

   \draw [-{stealth[scale=2.5]}] (0.0,-0.3) -- (0.0,3.8) node[above] {E} node[]{} ;
   \node[text width=0cm] at (\shiftA,-0.3) {El.};
   \node[text width=0cm] at (\shiftC,-0.3) {Ph.};

   \draw [line width=0.3mm] (\shiftA-\width,\ElC)  node[anchor=west,left]{$\ket{\Gamma_C,0}$}   -- (\shiftA+\width,\ElC)   ;
   \draw [line width=0.3mm] (\shiftA-\width,\ElB)  node[anchor=west,left]{$\ket{\Gamma_B,0}$}-- (\shiftA+\width,\ElB)  ;
   \draw [line width=0.3mm] (\shiftA-\width,\ElA)  node[anchor=west,left]{$\ket{\Gamma_A,0}$}   -- (\shiftA+\width,\ElA)   ;

   \draw [line width=0.3mm] (\shiftB-\width,\PoD)   -- node[above]{$\ket{P_+}$}  (\shiftB+\width,\PoD)   ;
   \draw [line width=0.3mm] (\shiftB-\width,\PoC)   -- node[anchor=west,left]{}  (\shiftB+\width,\PoC)   ;
   \draw [line width=0.3mm] (\shiftB-\width,\PoB)   -- node[above]{$\ket{P_-}$}  (\shiftB+\width,\PoB)  ;
   \draw [line width=0.3mm] (\shiftB-\width,\PoA)   -- node[anchor=west,left]{}  (\shiftB+\width,\PoA)   ;

   \draw [line width=0.3mm] (\shiftC-\width,\PhA)  -- (\shiftC+\width,\PhA)  node[right]{$\ket{\Gamma_A,1}$}    ;

   \draw [line width=0.3mm, gray, dotted] (\shiftA+\width,\ElC) -- (\shiftB-\width,\PoC)   ;
   \draw [line width=0.3mm, gray, dotted] (\shiftA+\width,\ElB) -- (\shiftB-\width,\PoD)  ;
   \draw [line width=0.3mm, gray, dotted] (\shiftA+\width,\ElB) -- (\shiftB-\width,\PoB)  ;
   \draw [line width=0.3mm, gray, dotted] (\shiftA+\width,\ElA) -- (\shiftB-\width,\PoA)   ;
   \draw [line width=0.3mm, gray, dotted] (\shiftB+\width,\PoD) -- (\shiftC-\width,\PhA) ;
   \draw [line width=0.3mm, gray, dotted] (\shiftB+\width,\PoB) -- (\shiftC-\width,\PhA) ;

\end{tikzpicture}
\end{minipage}
\caption{Exemplary energy landscape for an electronic three level system in presence of a single photonic mode.
The irreducible representations of the electronic ground state is $\Gamma_A$ and for the two excited states $\Gamma_B$ and $\Gamma_C$.
The irreducible representation of the photon is $\Gamma_\text{ph}$.
For the depicted system $\Gamma_A \otimes \Gamma_\text{ph} = \Gamma_B \ne \Gamma_C$ }
\label{fig:polariton1}
\end{figure}
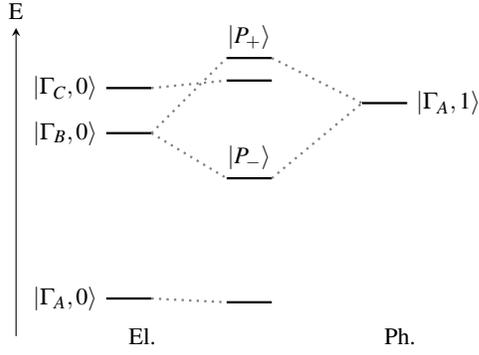

An exemplary energy diagram for the formation of polaritons of a three level system can be found in Fig. \ref{fig:polariton1}.
Note that for each electronic state $\ket{\Gamma_\text{el},0}$, there exists one photonic state $\ket{\Gamma_\text{el},1}$ shifted in energy by the cavity frequency $\omega_\alpha$.
In Fig. \ref{fig:polariton1} only the photonic state $\ket{\Gamma_A,1}$ is depicted, while in principle also the photonic states $\ket{\Gamma_B,1}$ and $\ket{\Gamma_C,1}$ exist, but are assumed to be  energetically too high for any significant coupling. 

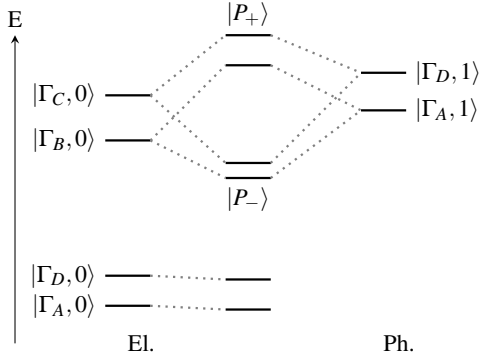
\begin{figure}
\begin{minipage}{1.0\linewidth}
\begin{tikzpicture}[]

   \def\shiftA{1.5}
   \def\shiftB{3.1}
   \def\shiftC{4.9}

   \def\width{0.3}

   \def\ElC{3.0}
   \def\ElB{2.4}
   \def\ElD{0.6}
   \def\ElA{0.2}

   \def\PoDb{3.8}
   \def\PoDa{3.4}
   \def\PoCb{3.0}
   \def\PoCa{2.4}
   \def\PoBb{2.1}
   \def\PoBa{1.9}
   \def\PoE{0.55}
   \def\PoA{0.15}

   \def\PhA{2.8}
   \def\PhB{3.3}

   \draw [-{stealth[scale=2.5]}] (0.0,-0.3) -- (0.0,3.8) node[above] {E} node[]{} ;
   \node[text width=0cm] at (\shiftA,-0.3) {El.};
   \node[text width=0cm] at (\shiftC,-0.3) {Ph.};

   \draw [line width=0.3mm] (\shiftA-\width,\ElC)  node[anchor=west,left]{$\ket{\Gamma_C,0}$}   -- (\shiftA+\width,\ElC)   ;
   \draw [line width=0.3mm] (\shiftA-\width,\ElB)  node[anchor=west,left]{$\ket{\Gamma_B,0}$}   -- (\shiftA+\width,\ElB)  ;
   \draw [line width=0.3mm] (\shiftA-\width,\ElD)  node[anchor=west,left]{$\ket{\Gamma_D,0}$}   -- (\shiftA+\width,\ElD)   ;
   \draw [line width=0.3mm] (\shiftA-\width,\ElA)  node[anchor=west,left]{$\ket{\Gamma_A,0}$}   -- (\shiftA+\width,\ElA)   ;

   \draw [line width=0.3mm] (\shiftB-\width,\PoDa)   -- node[above]{}  (\shiftB+\width,\PoDa)   ;
   \draw [line width=0.3mm] (\shiftB-\width,\PoDb)   -- node[above]{$\ket{P_+}$}             (\shiftB+\width,\PoDb)   ;
   \draw [line width=0.3mm] (\shiftB-\width,\PoBb)   -- node[below]{}             (\shiftB+\width,\PoBb)  ;
   \draw [line width=0.3mm] (\shiftB-\width,\PoBa)   -- node[below]{$\ket{P_-}$}  (\shiftB+\width,\PoBa)  ;
   \draw [line width=0.3mm] (\shiftB-\width,\PoE)   -- node[anchor=west,left]{}   (\shiftB+\width,\PoE)   ;
   \draw [line width=0.3mm] (\shiftB-\width,\PoA)   -- node[anchor=west,left]{}   (\shiftB+\width,\PoA)   ;

   \draw [line width=0.3mm] (\shiftC-\width,\PhA)  -- (\shiftC+\width,\PhA)  node[right]{$\ket{\Gamma_A,1}$}    ;
   \draw [line width=0.3mm] (\shiftC-\width,\PhB)  -- (\shiftC+\width,\PhB)  node[right]{$\ket{\Gamma_D,1}$}    ;

   \draw [line width=0.3mm, gray, dotted] (\shiftA+\width,\ElB) -- (\shiftB-\width,\PoDa)  ;
   \draw [line width=0.3mm, gray, dotted] (\shiftA+\width,\ElB) -- (\shiftB-\width,\PoBa)  ;
   \draw [line width=0.3mm, gray, dotted] (\shiftB+\width,\PoDa) -- (\shiftC-\width,\PhA) ;
   \draw [line width=0.3mm, gray, dotted] (\shiftB+\width,\PoBa) -- (\shiftC-\width,\PhA) ;
   \draw [line width=0.3mm, gray, dotted] (\shiftA+\width,\ElC) -- (\shiftB-\width,\PoDb)  ;
   \draw [line width=0.3mm, gray, dotted] (\shiftA+\width,\ElC) -- (\shiftB-\width,\PoBb)  ;
   \draw [line width=0.3mm, gray, dotted] (\shiftB+\width,\PoDb) -- (\shiftC-\width,\PhB) ;
   \draw [line width=0.3mm, gray, dotted] (\shiftB+\width,\PoBb) -- (\shiftC-\width,\PhB) ;

   \draw [line width=0.3mm, gray, dotted] (\shiftA+\width,\ElD) -- (\shiftB-\width,\PoE)   ;
   \draw [line width=0.3mm, gray, dotted] (\shiftA+\width,\ElA) -- (\shiftB-\width,\PoA)   ;

\end{tikzpicture}
\end{minipage}
\caption{Exemplary energy landscape for an electronic four level system in presence of a single photonic mode.
The irreducible representations of the electronic ground state is $\Gamma_A$ and for the three excited states $\Gamma_B$, $\Gamma_C$ and $\Gamma_D$.
The irreducible representation of the photon is $\Gamma_\text{ph}$.
For the depicted system $\Gamma_D \otimes \Gamma_\text{ph} = \Gamma_C \ne \Gamma_B$ and $\Gamma_A \ne \Gamma_D$.}
\label{fig:polariton2}
\end{figure}

A more complicated system is depicted in Fig. \ref{fig:polariton2} which leads to two sets of polaritons since  another electronic state $\ket{\Gamma_D,0}$ is energetically close to the ground state.
From this state the photonic state $\ket{\Gamma_D,1}$ can be formed which is of same symmetry as the state $\ket{\Gamma_C,0}$ and forms a second pair of upper and lower polaritons. 
In such a case with one cavity mode multiple polaritons are generated. 
Also, similar to the case of a molecule in a magnetic field, the symmetry of the system and the resulting energy landscape depends on the orientation of the molecule in the cavity which influences the formation of upper and lower polaritons.
To resemble an experimental setup an entire ensemble of different orientations would need to be considered.

\section{Results and Discussion}
\label{sec:calculations}
All calculations presented in this work were carried out using a developer's version of the CFOUR\cite{cfour,matthews2020coupled} and Qcumbre\cite{qcumbre,hampe2017equation} program packages for the HF and CC steps, respectively.
The symmetry effects inside a cavity were investigated for the H$_2$ and H$_2^-$ molecules.  
The energy was tracked with respect to the bond distance in increments of $0.12 \;a_0$ for a parallel and perpendicular orientation of the polarization vector, respectively.
All calculations were performed in a cc-pVTZ basis\cite{dunning1989gaussian} and were converged to $10^{-8} E_\text{h}$. 
In the parallel orientation  $D_{\infty h}$ symmetry is retained, in the perpendicular orientation the system has $D_{2h}$ symmetry. 

\begin{figure}[]
   \centering
   \includegraphics[width=\linewidth]{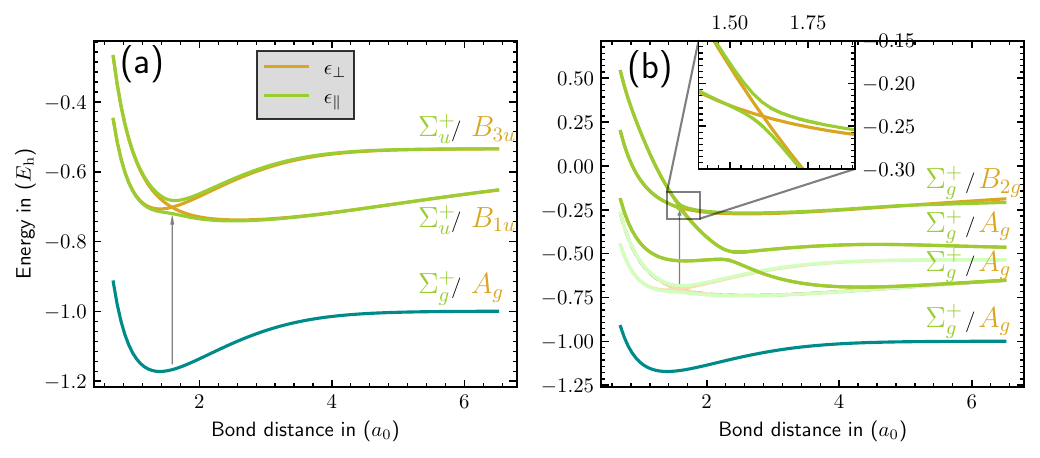}
   \caption{Low-lying singlet states of H$_2$ in a cavity with the polarization vector aligned parallel (green) and perpendicular (orange) with respect to the molecular axis.
         Excited states of \textit{ungerade} and \textit {gerade} parity on left (a) and right-hand side (b), respectively. 
            In b) in faded colors also states of \textit {ungerade} parity are shown.
            The cavity frequency was set to $0.466 E_\text{h}$ and the coupling strength to $0.03 \text{a.u.}$
            The arrows indicate where a polaritonic coupling occurs and from which electronic state the photonic excitation occurs.  
            }
   \label{fig:h2_sing}
\end{figure}

The lowest singlet states of the H$_2$ molecule are depicted in Fig.  \ref{fig:h2_sing}, where  Fig. \ref{fig:h2_sing}a shows the ground state plus the two lowest excited states of \textit {ungerade} parity and Fig.  \ref{fig:h2_sing}b shows the ground state together with the three lowest excited states of \textit {gerade} parity. 
In Fig. \ref{fig:h2_sing}b, in addition to the \textit{gerade} states also the states of \textit {ungerade} parity are depicted (in faded colors) to better clarify the coupling mechanisms.
Different notations for the total irreducible representation of the state as well as the contributions from the electronic and photonic parts are listed in table \ref{tab:symtable}.
Fig.  \ref{fig:h2_sing}a shows that no splitting in upper and lower polariton occurs for the perpendicular orientation of the polarization vector as the electronically excited state $\ket{B_{1u},0}$ and the photonic excited state $\ket{A_g,1}$ are characterized by different irreducible representations.
Note that the $\ket{A_g,1}$ state is the ground-state shifted by the frequency of the cavity $\omega$ which  therefore has a purely photonic character.
For the parallel orientation of the polarization vector, however, 
$\ket{\Sigma_u^+,0}$ and $\ket{\Sigma_g^+,1}$ have to belong to the same total irreducible representation $\Sigma_u^+$ and a coupling resulting in an upper and lower polariton is observed.

\begin{table*}[]
    \centering
    \begin{tabular}{c|l|l}
       state         & \multicolumn{1}{c|}{parallel}  & \multicolumn{1}{c}{perpendicular} \\
       \hline
       \hline
       $\ket{0}$  & $\ket{\Sigma_g^+} = \ket{\Sigma_g^+,0} = \ket{\Sigma_g^+, \Sigma_g^+} $   &  $\ket{A_{g\;}} = \ket{A_{g\;},0} = \ket{A_g, A_g} $ \\
       $\ket{1}$ & $\ket{\Sigma_u^+} = \ket{\Sigma_u^+,0} = \ket{\Sigma_u^+,\Sigma_g^+}$ &$\ket{B_{1u}} = \ket{B_{1u},0} = \ket{B_{1u},A_g}$ \\ 
       $\ket{2}$ &  $\ket{\Sigma_u^+} = \ket{\Sigma_g^+, 1} = \ket{\Sigma_g^+,\Sigma_u^+}$ & $\ket{B_{3u}} = \ket{A_g,1} = \ket{A_g, B_{3u}} $\\
       \hline
        $\ket{3}$,$\ket{4}$ & $\ket{\Sigma_g^+} = \ket{\Sigma_g^+,0} = \ket{\Sigma_g^+,\Sigma_g^+}$ & $\ket{A_g} = \ket{A_g,0} = \ket{A_g,A_g}$ \\
        $\ket{5}$ & $\ket{\Sigma_g^+} = \ket{\Sigma_u^+,1} = \ket{\Sigma_u^+,\Sigma_u^+}$ & $\ket{B_{2g}} = \ket{B_{1u},1} = \ket{B_{1u}, B_{3u}}$ \\
        \hline
        $\ket{6}$ & $\ket{\Sigma_g^+} = \ket{\Sigma_g^+,2} = \ket{\Sigma_g^+,\Sigma_g^+} $ & $\ket{A_g} = \ket{A_g,2} = \ket{A_g,A_g}$ \\
        \hline
        \hline
    \end{tabular}
    \caption{Different notations for the polaritonic states of $^1$H$_2$ in parallel as well as perpendicular orientation.}
    \label{tab:symtable}
\end{table*}

\begin{figure}[]
   \centering
   \includegraphics[width=\linewidth]{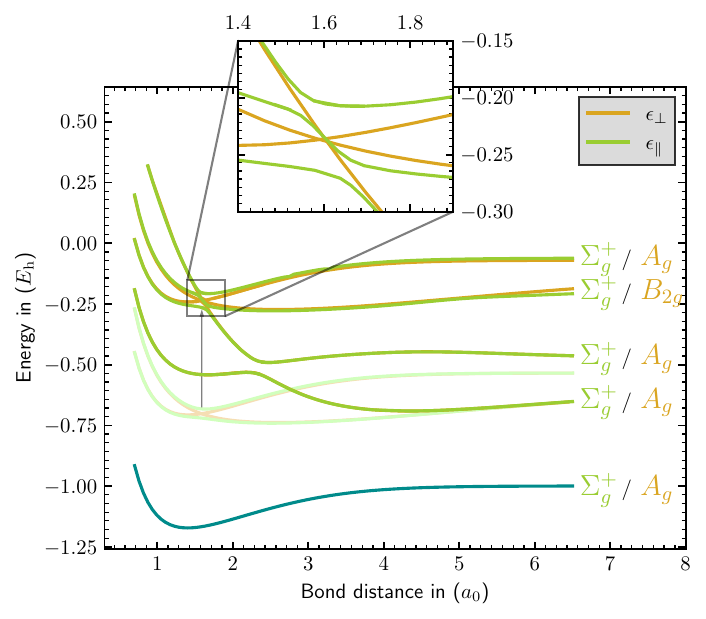}
   \caption{Singlet states of  H$_2$ in a cavity with the polarization vector aligned parallel and perpendicular with respect to the molecular axis.
            States of \textit {gerade} parity are found on the left hand side and states of \textit {ungerade} parity on the right hand side (in faded colors also states of \textit {gerade} parity are shown).
            The cavity frequency was set to $0.466 E_\text{h}$ and the coupling strength to $0.03 \text{a.u.}$
            The arrows indicate where a polaritonic coupling occurs and from which
            electronic state the photonic excitation occurs.}
   \label{fig:h2_sing_g2}
\end{figure}

\begin{figure}[]
   \centering
   \includegraphics[width=\linewidth]{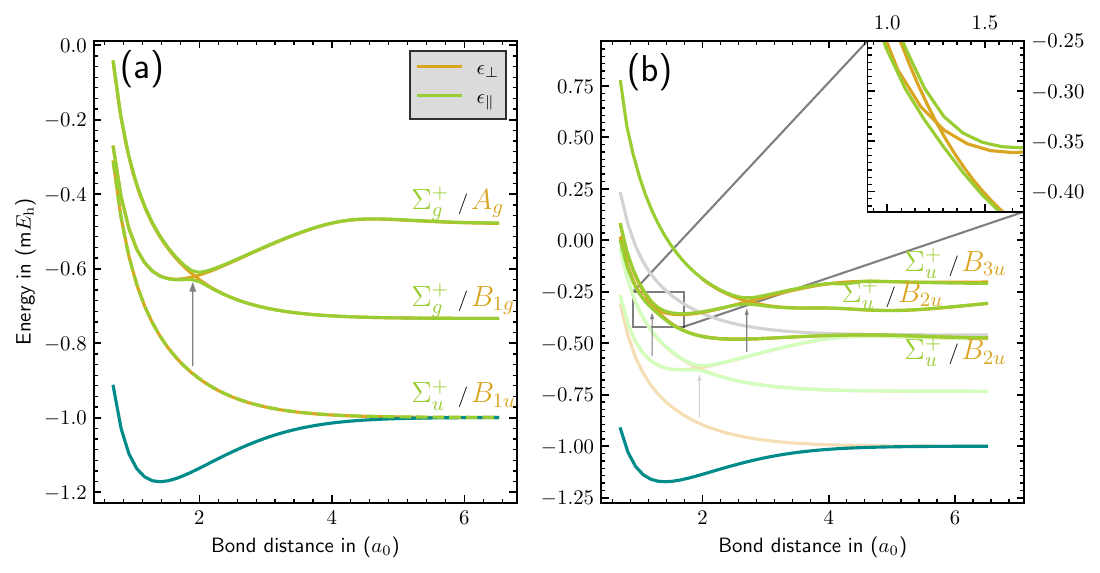}
   \caption{ Triplet states of the H$_2$ molecule in a cavity with the polarization vector aligned parallel and perpendicular with respect to the molecular axis.
            States of \textit {gerade} parity are found on the left hand side and states of \textit {ungerade} parity on the right hand side (in faded colors also states of \textit {gerade} parity.
            The cavity frequency was set to $0.466 E_\text{h}$ and the coupling strength to $0.03 \text{a.u.}$
            The arrows indicate where a polaritonic coupling in lower and upper polariton occurs and which state is included.}
   \label{fig:h2_trip}
\end{figure}

\begin{figure}[]
   \centering
   \includegraphics[width=\linewidth]{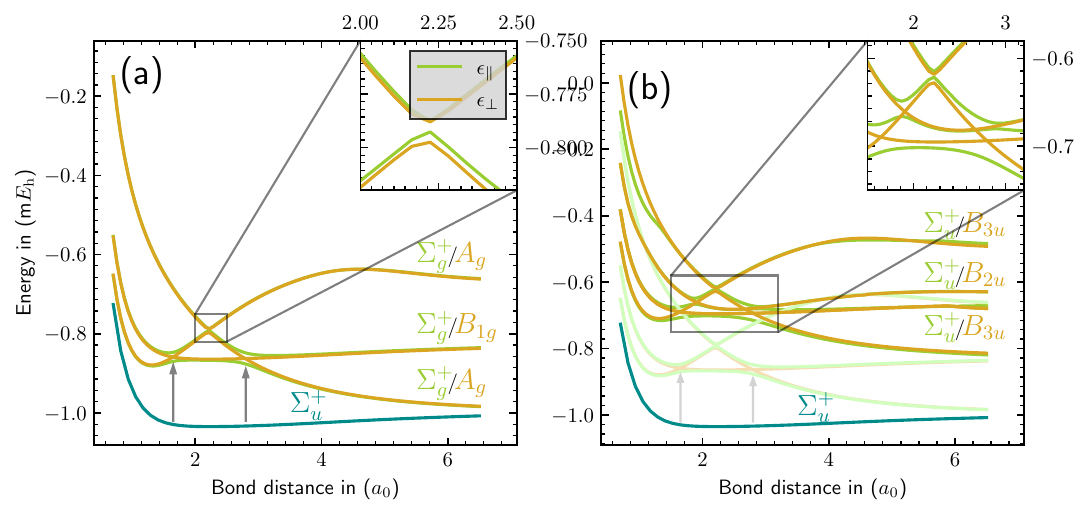}
   \caption{ Doublet states of  H$_2^-$  in a cavity with the polarization vector aligned parallel and perpendicular with respect to the molecular axis.
            States of \textit {ungerade} parity are found on the left hand side and states of \textit {gerade} parity on the right hand side (in faded colors also states of \textit {ungerade} parity.
            The cavity frequency was set to $0.466 E_\text{h}$ and the coupling strength to $0.03 \text{a.u.}$
            The arrows indicate where a polaritonic coupling in lower and upper polariton occurs and which state is included.}
   \label{fig:h2_doub}
\end{figure}

Fig.  \ref{fig:h2_sing}b shows the coupling mechanism for the states of \textit {gerade} parity. 
In the perpendicular orientation no coupling in upper and lower polariton can be observed as the two lowest-lying excited electronic states $\ket{A_g,0}$ are of $A_g$ symmetry while the photonic state $\ket{B_{1u},1}$ is of $B_{2g}$ symmetry.
The photonic state is in fact the $\ket{B_{1u},0}$ state from Fig. \ref{fig:h2_sing}a shifted by the frequency of the cavity.
For the parallel orientation the coupling in upper and lower polariton occurs as all three states are of $\Sigma_g^+$ symmetry. Those are the two electronic states $\ket{\Sigma_g^+,0}$ and one photonic state $\ket{\Sigma_u^+, 1}$.
The arrow again indicates where the polariton is formed and which state is participating ($\ket{\Sigma_u^+,0}$).
However, note that an additional state of type $\ket{\Sigma_g^+, 2}$ would be expected which is absent in Fig. \ref{fig:h2_sing}b as it is not represented in the CCSD-1-SD truncation scheme when starting from the $\ket{\Sigma_g^+ , 0}$ state as the CC reference.
The description of the $\ket{\Sigma_g^+, 2}$  state would require doubly excited photonic states which are described to a large part by the $R_{0,2}$ operator.  
The CCSD-12-SD truncation scheme is indeed able to also describe photonic doubly excited states as seen in Fig. \ref{fig:h2_sing_g2}.
Here, the missing $\ket{\Sigma_g^+,2}$ state appears in close vicinity to the avoided crossing. 
Note, however, that only states differing by at most one photon can couple.
The interaction appearing for the parallel orientation can hence be viewed as two separate avoided crossings, one appearing between the electronic state $\ket{\Sigma_g^+,0}$ and the $\ket{\Sigma_u^+,1}$ state and the second between the photonic doubly excited ground-state $\ket{\Sigma_g^+,2}$ and the state $\ket{\Sigma_u^+,1}$, respectively.
For the perpendicular orientation, no polaritonic splitting is observed within the investigated energy range. \\
In Fig. \ref{fig:h2_trip} the triplet states of the H$_2$ molecule are shown. 
Again, Fig.  \ref{fig:h2_trip}a shows the states of \textit{gerade} and Fig.  \ref{fig:h2_trip}b the states of \textit{ungerade} parity.
As for the singlet, Fig.  \ref{fig:h2_trip}a also shows in faded colors the lowest triplet state of \textit {ungerade} parity. 
Also here the upper and lower polariton is only formed for the parallel field orientation for the states and cavity frequency considered.
In Fig.  \ref{fig:h2_trip}b three states of \textit {ungerade} parity can be found.
Two of them are electronic states of the type $\ket{\Sigma_u^+,0}$ and one is a photonic state of type $\ket{\Sigma_g^+,1}$. 
Since the photonic state crosses both electronic states two avoided crossings are generated for the parallel field orientation (indicated by the arrows). 
However, note that actually a fourth state would be expected, appearing energetically between the two electronic states of $\ket{\Sigma_u^+,0}$ symmetry. 
It corresponds to a photonic double excitation of the lowest triplet state ($\ket{\Sigma_u^+,2}$). However, to describe this state, the inclusion of the $R_{1,2}$ operator would be required which is beyond the scope of the present investigation. 
This case overall shows polaritons of different symmetries similar to the situation discussed in chapter \ref{sec:symexc}.
\\Finally, the H$_2^-$ anion is investigated, see Fig. \ref{fig:h2_doub}, based on our spin-unrestricted QED-EOM-CC implementation.
Typically, open-shell species do not have totally symmetric ground states, leading to more complicated coupling mechanisms inside a cavity. 
Fig.  \ref{fig:h2_doub}a shows  the $\Sigma_u^+$ reference state, two electronically excited states $\ket{\Sigma_g^+,0}$,  and one photonic state $\ket{\Sigma_u^+,1}$.
The formation of the upper and lower polariton is found at about $1.6 a_0$ and $2.8 a_0$ with the lower electronic state. 
Note that there is also a purely electronic avoided crossing around $2.1 a_0$ in between the two states. 
States of \textit {ungerade} parity can be found in Fig. \ref{fig:h2_doub}b showing a rather complicated coupling mechanism of four states.
One of them is purely electronic, $\ket{\Sigma_u^+,0}$, two are single electronic single photon excitations $\ket{\Sigma_g^+,1}$ and one is a double photonic excitation from the ground-state $\ket{\Sigma_u^+,2}$.
For the parallel orientation all states couple and several avoided crossings are observed. 
(Note that the $\ket{\Sigma_u^+,0}$ and $\ket{\Sigma_u^+,2}$ states are not directly coupled since they differ by more than one photon.)
Certainly, it becomes clear that even for relatively simple systems like H$_2$ and H$_2^-$ there is already quite noticeable complexity in the formation of polaritonic states. In the electronic strong-coupling regime, hence, an accurate description of the involved states is inevitable.

\section{Conclusion}
We have generalized the well established diagrammatic notation for CC theory to derive the QED-CC and QED-EOM-CC equations in an intuitive manner.
The presented diagrammatic formalism is applicable for any truncation schemes of the cluster operator and has here been used to derive the (EOM-)CCSD-1-SD and  the (EOM-)CCSD-12-SD truncations.
For the interpretation of diagrams including photonic indices only few additional rules are needed which are extensions to the rules for electronic CC diagrams.
We discussed an adaption of a CIS guess for the QED-EOM-CC context  
by projecting also on photonic excited states.

For symmetric molecules, the symmetry handling and targeted calculation of individual states is particularly interesting if one is interested in the states that show polaritonic coupling.
We discussed exemplary polaritonic splitting schemes thereby investigating H$_2$ and H$_2^-$. 
Both show a surprisingly complicated energetic landscapes in a cavity with the polarization vector parallel to the molecular axis.

\begin{acknowledgments}
This paper is dedicated to Rodney Bartlett. The authors acknowledge his numerous contributions to CC theory not least the classic book on diagrams in CC and EOM-CC theory  which the research group in Saarbrücken is benefiting from to this day. This work was funded by the Deutsche Forschungsgemeinschaft (DFG, German Research Foundation) – Project-ID 429529648 – TRR 306 QuCoLiMa (“Quantum Cooperativity of Light and Matter’’).
\end{acknowledgments}

\appendix

\section{Diagrammatic expression for the S amplitude equations}

\begin{figure*}
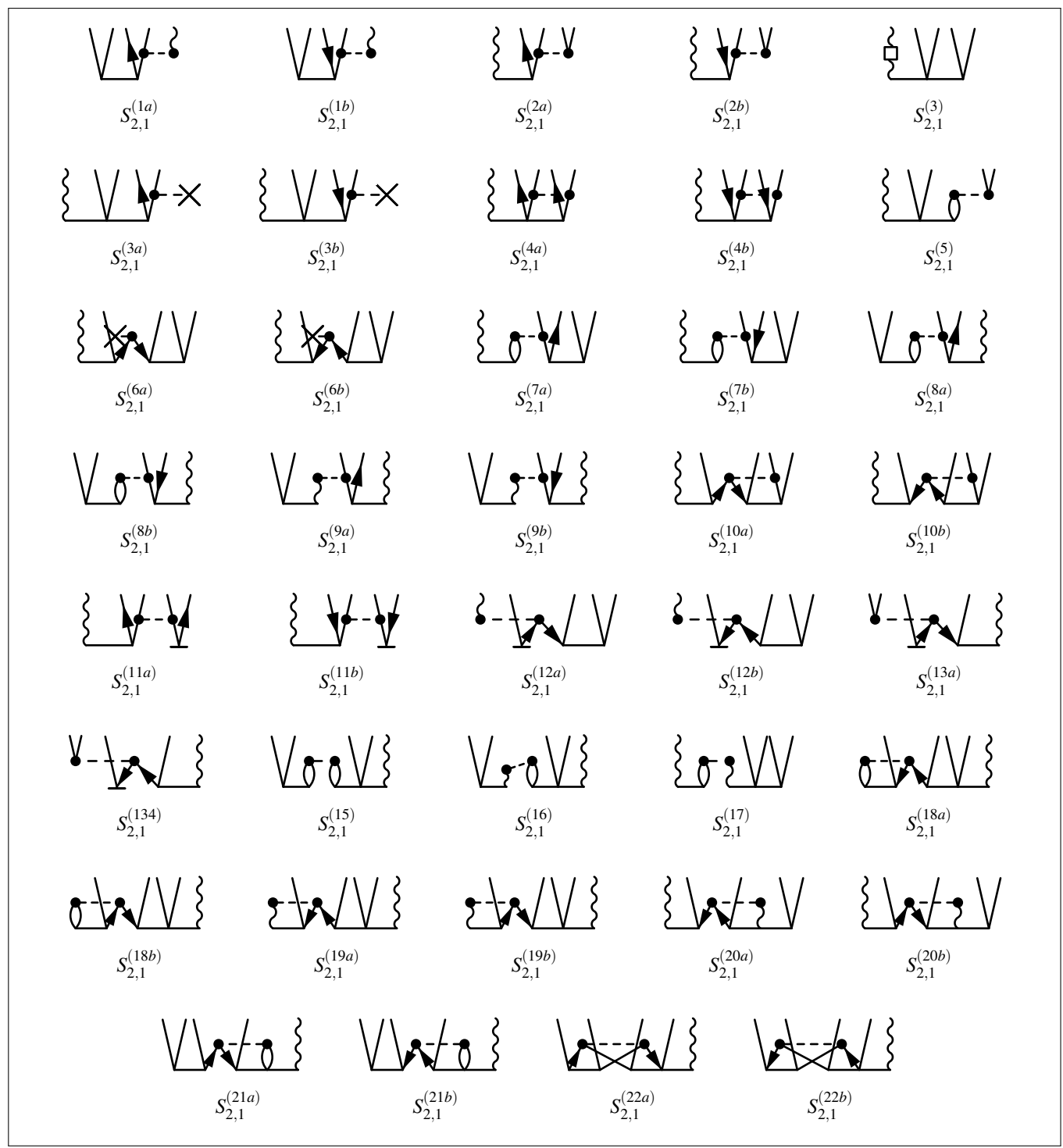

\begin{framed}
\include{include/S21-amplitudes.tex}
\end{framed}
\caption{Diagrams to determine the  $S_2^1$ amplitudes.}
\label{fig:S21-1}
\end{figure*}
\begin{figure*}
\begin{framed}
\include{include/S21-amplitudes2.tex}
\end{framed}
\caption{Diagrams to determine the  $S_2^1$ amplitudes.}
\label{fig:S21-2}
\end{figure*}

\nocite{*}
\bibliography{lit}

\end{document}